\documentclass[aps,onecolumn]{revtex4}
\usepackage{amsmath}
\usepackage{latexsym}
\usepackage{float}
\usepackage{amssymb}
\usepackage{graphicx}
\usepackage{epsfig}
\usepackage{psfrag}
\usepackage{color}
\usepackage{multirow}
\usepackage{ulem}

\begin{document}

\title{Hierarchical modeling of polystyrene melts: From soft blobs to atomistic resolution}

\author{Guojie Zhang}
\affiliation{Max Planck Institute for Polymer Research, Ackermannweg 10, 55128 Mainz, Germany}
\affiliation{
School of Chemistry and Chemical Engineering, Institute for Systems Rheology, Advanced Institute of Engineering Science for Intelligent Manufacturing, Guangzhou University, 510006 Guangzhou, China. E-mail address: guojie.zhang@gzhu.edu.cn}

\author{Anthony Chazirakis}
\affiliation{Department of Mathematics and Applied Mathematics, University of Crete, GR-71409 Heraklion, Crete, Greece}
\affiliation{Institute of Applied and Computational Mathematics, IACM/FORTH, Heraklion, Greece}

\author{Vagelis A. Harmandaris}
\altaffiliation{Corresponding author e-mail: harman@uoc.gr}
\affiliation{Department of Mathematics and Applied Mathematics, University of Crete, GR-71409 Heraklion, Crete, Greece}
\affiliation{Institute of Applied and Computational Mathematics, IACM/FORTH, Heraklion, Greece}

\author{Torsten Stuehn}
\affiliation{Max Planck Institute for Polymer Research, Ackermannweg 10, 55128 Mainz, Germany}

\author{Kostas Ch. Daoulas}
\altaffiliation{Corresponding author e-mail: daoulas@mpip-mainz.mpg.de}
\affiliation{Max Planck Institute for Polymer Research, Ackermannweg 10, 55128 Mainz, Germany}

\author{Kurt Kremer}
\altaffiliation{Corresponding author e-mail: kremer@mpip-mainz.mpg.de}
\affiliation{Max Planck Institute for Polymer Research, Ackermannweg 10, 55128 Mainz, Germany}

\begin{abstract}
We demonstrate that hierarchical backmapping strategies incorporating generic blob-based models can equilibrate melts of high-molecular-weight polymers, described with chemically specific, atomistic, models. 
The central idea behind these strategies, is first to represent polymers by chains of large soft blobs (spheres) and efficiently equilibrate the melt on mesoscopic scale. Then, the degrees of freedom of more detailed models are reinserted step by step. The procedure terminates when the atomistic description is reached. Reinsertions are feasible computationally because the fine-grained melt must be re-equilibrated only locally. 
To develop the method, we choose a polymer with sufficient complexity. We consider polystyrene (PS), characterized by stereochemistry and bulky side groups. Our backmapping strategy bridges mesoscopic and atomistic scales by incorporating a blob-based, a moderately CG, and a united-atom model of PS. We demonstrate that the generic blob-based model can be parameterized  to reproduce the mesoscale properties of a specific polymer -- here PS. The moderately CG model captures stereochemistry. 
To perform backmapping we improve and adjust several fine-graining techniques. We prove equilibration of backmapped PS melts by comparing their structural and conformational properties with reference data from smaller systems, equilibrated with less efficient methods. 
\end{abstract}

\maketitle


\let\clearpage\relax

\section{Introduction}
\label{sec:introduction}

A molecular structure based understanding of properties of technically or experimentally relevant polymer melts or dense solutions still poses significant scientific 
challenges. Computational studies of structure-process-property relationships in polymer liquids often require the consideration of atomistic details, 
while at the same time many fundamental scientific questions and important technological applications can be addressed only for high molecular-weight (MW) polymers. 
These simultaneous requirements create major challenges at high polymer concentrations, e.g. concentrated solutions and melts. For typical polymerization degrees 
in industrial applications, the average spatial extension of random-walk-like chains lies~\cite{GrosbergBook} between $10$ and $100$\;nm. In concentrated polymer liquids 
these fractal ``threads'' strongly interdigitate. Therefore, samples with dimensions only a few times larger than the average chain size contain hundreds or even thousands of polymers. 
This easily corresponds to many millions of atomistic degrees of freedom. The overlapping polymers are strongly entangled~\cite{Doi} and their relaxation times are 
prohibitively long. Their equilibration with atomistic molecular dynamics (MD) is unfeasible even when using massively parallel simulations. Therefore approaches, 
circumventing the slow relaxation path of physical dynamics are of significant interest. They are an indispensable starting part to simulation studies of atomistic long-chain polymer melts. Various strategies are available, including advanced connectivity altering Monte Carlo (MC) algorithms~\cite{Vlasis1999,Doros2007} and methods based on hierarchical multiscale modeling.

Hierarchical multiscale modeling involves multiple scales of description~\cite{Tschoep1,Tschoep2,Kremer2002,Voth2008,VothBook,Peter,Noid2013,HarmandPS} and offers 
a powerful concept for tackling large system sizes and protracted equilibration times. The central idea is to 
benefit from scale separation: in polymers long-wavelength behavior often follows universal laws~\cite{deGennes} that incorporate chemistry-specific information into a few parameters. Therefore, one can initially eliminate microscopic features hampering equilibration and prepare samples reproducing only mesoscopic conformational and structural properties. These samples serve as ``blueprints'' for reinserting the missing microscopic features without disturbing long-wavelength properties. The reinsertion is computationally feasible because only local re-equilibration is required. Two major strategies realize this idea: configuration assembly and hierarchical backmapping.

Configuration assembly~\cite{Auhl,Moreira,HarmandPS,Carbone,Sliozberg2016} algorithms construct the initial sample by putting together polymer chains under fixed average density. Implementations vary in details, but in all cases chains are generated to reproduce prescribed distributions of conformations. If the ensemble is generated as an ideal gas of chains, important long-wavelength structural properties, such as the correlation hole effect~\cite{deGennes}, are missing. Therefore, one must reduce the large density fluctuations of the ideal gas of chains and introduce some interchain correlations through chain packing schemes~\cite{Auhl,Moreira}. Subsequently local conformations and liquid packing are recovered through a ''push off'' procedure, which gradually~\cite{Auhl} reinserts the microscopic excluded volume into the ensemble. The strategy has been successful with generic~\cite{Auhl,Moreira} and chemistry specific~\cite{HarmandPS,Carbone,Sliozberg2016} microscopic models. 
However, the postulative construction of starting configurations is a drawback. The assumption that polymers in melts have ideal random-walk-like conformations is only approximately true~\cite{Strasbourg,Morse}. Examples of deviations from ideal random-walk statistics are found in the power-law decay of bond-bond correlations~\cite{Strasbourg} and statistics of polymer knots~\cite{Virnau}. Predicting conformations for polymers with complex architecture (star-like or branched) and systems that are inhomogeneous or multicomponent is even less straightforward. Furthermore, the computational costs at the stage of chain packing increase with molecular weight.

Hierarchical backmapping is a more general approach, taking advantage of a general concept originating from renormalization group theory in critical phenomena. The material is described at several ``nested'' length scales, introducing a sequence of coarse-grained (CG) models. The sequence is terminated by the microscopic model. Because the sample is equilibrated at the largest length scale with standard simulations (handling the crudest CG model), long-wavelength properties are not postulated but follow from a rigorous statistical-mechanical framework. The molecular details resolved by the next model are reinserted through local sampling of configuration space. Repeating the procedure, one descends the hierarchy of models, step by step, until the microscopic description is reached. The efficiency increases significantly~\cite{Steinmueller,Guojie2, Ozog,Svaneborg,Sliozberg2016,ParkerRottler} when the hierarchy includes soft models, i.e. models where the strength of non-bonded interactions is comparable to the thermal energy. 

Hierarchies of CG models can be constructed taking advantage of universalities in polymer behavior, using the classical concept~\cite{deGennes} of blobs. The decimation of the microscopic degrees of freedom is what resembles renormalization group theory:~\cite{deGennes,FreedBook} $N_{\rm b}$ monomers of a microscopically-resolved subchain are lumped into a single soft blob (sphere), so that polymers are represented by chains of blobs. 
Varying $N_{\rm b}$ generates a family of models with different resolutions. In a polymer melt, a key quantity controlling~\cite{Strasbourg,Morse,MuelBinder,Guenza3} conformations and liquid structure on the scale of blobs is given by $\sqrt{\bar{N}_{\rm b}} = \rho R_{\rm b}^3/N_{\rm b}\sim \sqrt{N_{\rm b}}$. $\bar{N}_{\rm b}$ is the invariant degree of polymerization of subchains, $R_{\rm b}$ is the root mean-square end-to-end distance of subchains, and $\rho$ is the number density of microscopic monomers. When the entire chain is considered, instead of a subchain, $\bar{N}_{\rm b}$ reduces to the invariant degree of polymerization of the melt, $\bar{N}$. 
Melts with the same $\bar{N}$ form a single class of materials which can be described~\cite{Guojie3} (in renormalized space) by a single blob-based model. This common model can be used to inter-convert~\cite{Guojie3} chemically different materials within the same $\bar{N}$-class. Increasing $\sqrt{\bar{N}_{\rm b}}$ has two important consequences: a) the conformations of subchains approach the Gaussian statistics of ideal random-walks and b) the correlation hole of blobs becomes more shallow, i.e. the intermolecular correlation function of blobs approaches unity. 
For both cases the small parameter controlling convergence is $1/\sqrt{\bar{N}_{\rm b}}$. Therefore, for large $\sqrt{\bar{N}_{\rm b}}$ the interactions in blob-based models can be approximated by generic expressions~\cite{Suter,Vettorel,Guojie1,Guojie3} inspired by the ideal random-walk limit. Deviations from the asymptotic Gaussian behavior are taken implicitly into account by ``renormalizing'' the parameters in the generic expressions. 

Hierarchical backmapping using blob-based models has been successful in equilibrating high-molecular-weight polymer melts~\cite{Guojie2,Guojie3} and blends~\cite{Takahiro} described with {\it generic} (bead-spring like) microscopic models. The equilibrated samples typically contained $10^3$ chains, comprised of a few thousands of beads. The method is not limited to these examples -- the computational costs of the procedure are not affected by chain length and are, roughly, proportional only to the volume of the system.~\cite{Guojie2} Yet, the question whether hierarchical strategies with blob-based models can equilibrate melts described with {\it chemically-specific} atomistic models remains, so far, open. Concerns regarding this point have been expressed in the literature~\cite{Sliozberg2016}. 


Here we address this question and demonstrate that hierarchical strategies with generic blob-based models can equilibrate melts of actual polymers, described with {\it chemically-specific} atomistic models. To illustrate this we develop a hierarchical backmapping method which equilibrates large atomistically-resolved samples of polystyrene (PS) melts. PS is a basic commodity material and is well-suited for method development: it is well studied experimentally and theoretically, and it is a sufficiently complex polymer, where molecules have tacticity and bulky units (benzene rings). 
To descent the hierarchy of scales, from coarse to atomistic, our backmapping strategy incorporates a blob-based,~\cite{Vettorel,Guojie1} a moderately~\cite{HarmandRev,VHKK09b} CG, and a united-atom~\cite{HarmandRev,CG-VH2010} (UA), model of PS. We demonstrate that a generic blob-based model~\cite{Vettorel,Guojie1,Guojie3} can describe the rather complex PS melt, accurately enough to allow backmapping. So far, chemistry-specific blob-based models have been developed for simpler polymers (e.g. polyethylene) using the machinery of Integral Equation theory.\cite{Guenza_2012a,Guenza_2012b} 
To perform backmapping within the hierarchy of chemically-specific models, we improve and adjust several reinsertion techniques.~\cite{Auhl,Moreira,HarmandPS} These methodological issues are also discussed in the paper.

\section{Hierarchy of models}
\label{sec:models}

Our hierarchical strategy for equilibrating large samples of high-molecular-weight polymer melts, described with atomistic detail, requires several models covering a broad range of length scales: from atomistic, to moderately CG up to mesoscopic one. In this section we present the models used for the hierarchical description of PS.

\begin{figure}[ht]
\includegraphics[width=0.45\textwidth]{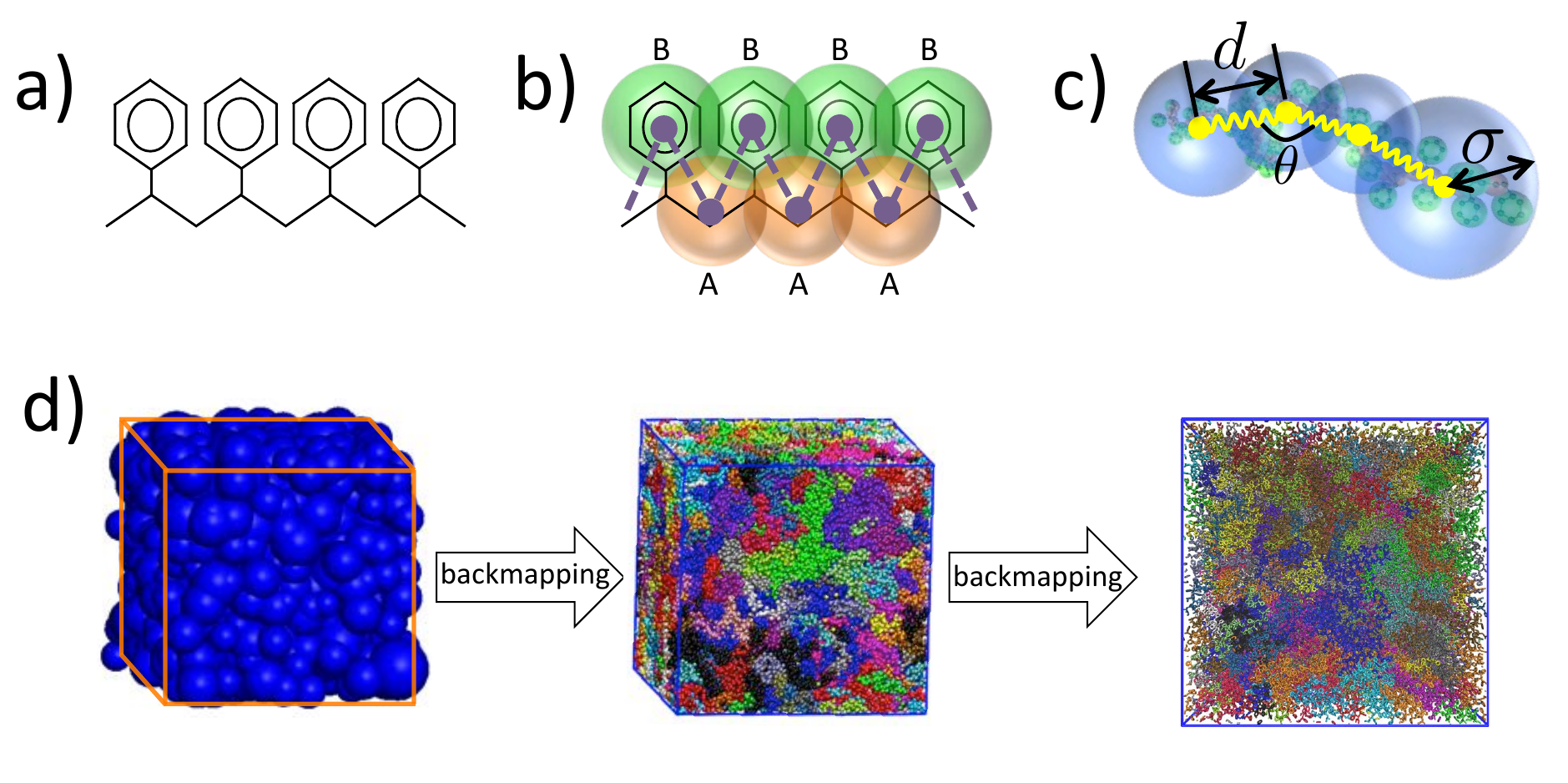}
\caption{PS models used during hierarchical backmapping: (a) United atom (UA) model; (b) moderately coarse-grained (CG) model. A-type beads (orange) represent a ${\rm CH_2}$ group and half of its
two neighboring $\rm CH$ groups along the aliphatic backbone, while B-type beads (green) describe the phenyl rings; and (c) Blob-based (BL) model constructed by grouping
a large number of beads of the moderately CG model into soft spheres. (d) Summary of basic steps of the hierarchical backmapping. An equilibrated blob-based
configuration of PS is first obtained from efficient Monte Carlo (MC) simulations and the degrees of freedom of the moderately CG model are reinserted into this configuration.
The equilibrated PS melt obtained after this reinsertion is further backmapped on the UA description. The snapshots show an equilibrated PS melt with 100 chains and 480 monomers
per chain at three different resolutions.}
\label{fig:hierarchical_modeling}
\end{figure}

\subsection{Atomistic model}
\label{model:AtomPS}
On atomistic level we describe PS through a united atom (UA) model based on the TraPPE-UA force field~\cite{TraPPE}. Details can be found in previous studies~\cite{HarmandRev,CG-VH2010}, so we mention here briefly that in this UA model each PS monomer is represented by groups of eight UAs (see Fig.~\ref{fig:hierarchical_modeling}(a)). 
The full interaction potential of atomistic PS involves bonded and non-bonded terms. Angular and torsional potentials are introduced for the aliphatic backbone, while keeping the lengths of the bonds fixed. Improper dihedral potentials are used to keep
the phenyl ring planar and to maintain the tetrahedral configuration around the sp3-hybridized carbon connecting the phenyl ring. Non-bonded interactions between UAs are captured by pairwise Lennard-Jones (LJ) potentials.

Despite the fact that we are using a UA model it is rather straightforward to obtain and simulate all-atom systems by adding hydrogens in the UA PS configurations. However, since the UA PS model has been extensively used and examined before, we choose to represent the atomistic PS chains in the UA description. 

\subsection{Moderately coarse-grained model}
\label{model:CGPS}
We employ a moderately (quantitative) CG model that has been derived~\cite{VH2006,HarmandPS} from the UA description presented in Sec~\ref{model:AtomPS}. The model and the procedure used to obtain the CG force field have been elaborated elsewhere.~\cite{HarmandPS,HarmandRev} 
In summary, each PS monomer is mapped on two effective beads, ``superatoms'', A and B (see Fig.~\ref{fig:hierarchical_modeling}(b)). Bead A represents the ${\rm CH_2}$ unit of a PS monomer and the half of each of the two neighboring $\rm CH$ groups along the chain backbone. Bead B stands for the phenyl ring. The sizes and masses of A and B superatoms are $\sigma_{\rm A}=4.1$ \AA, $\sigma_{\rm B}=5.2$ \AA, $m_{\rm A}=27$ amu, $m_{\rm B}=77$ amu, respectively. Such a model is capable of providing quantitative predictions about both static and dynamic properties of PS systems.
In addition, for the moderately CG model we define: (a) a characteristic length scale as the averaged bead size, i.e., $\sigma_{\rm CG}=(\sigma_{\rm A}+\sigma_{\rm B})/2 = 4.65$ \AA, and (b) a characteristic time scale, which is defined by $\tau = \sqrt{m_{\rm A} \sigma_{\rm CG}^2/k_{\rm B}T}$  ($k_{\rm B}T$ is the thermal energy). 

This CG scheme can describe tacticity of PS chains without introducing side groups, by classifying the B beads into four subtypes.  In total, the bonded part of the CG force field 
includes one bond, but four ``alternative'' angular and dihedral interaction potentials. The specific sequence of potentials chosen for angular and dihedral interactions along the backbone of a moderately CG chain depends on the tacticity of the atomistic PS molecule it represents. 

Nonbonded interactions are described through pair LJ-like, $n$-$m$ potentials $U_{\alpha,\beta}(r)$, where $r$ is the distance between two interacting superatoms. The powers and parameters used in $U_{\alpha,\beta}(r)$ depend on the type $\alpha$ and $\beta$ of the 
interacting superatoms, i.e. a bead can be $A$ type or belong to one of the four $B$ types. 
Non-bonded interactions are deactivated for those beads that belong to the same chain and are closer neighbours than $(\rm 1,5)$. 
The details and the parameters of the force-field can be found elsewhere.~\cite{VH2006,HarmandPS} 

The accuracy of the moderately CG model in predicting {\it quantitatively} the properties of PS melts 
has been verified through extensive investigations.~\cite{VH2006,HarmandPS} For the purposes of our backmapping scheme, the fine structure of the CG model presents significant advantages because it facilitates the subsequent reinsertion of the degrees of freedom of the UA description.

\subsection{Blob-based coarse-grained model}
\label{model:blobPS}
\subsubsection{Model description}
\label{model:blobPSdef}

Earlier studies~\cite{Vettorel,Guojie1,Guojie2,Guojie3,Takahiro} introduced blob-based models for polymer melts described on microscopic level through generic (bead-spring) models.~\cite{KremerGrest} Mapping the chemistry-specific moderately CG model of PS on a blob-based description is similar to these cases. 

A moderately CG polystyrene molecule with $N_{\rm CG}$ superatoms is represented by a sequence of $N_{\rm BL} = N_{\rm CG}/N_{\rm b}$ spheres (blobs). As illustrated in Fig.~\ref{fig:hierarchical_modeling}(c), each of them stands for a PS subchain with $N_{\rm b}$ superatoms, so that the choice of $N_{\rm b}$ determines the resolution of the drastically CG description. The radius $\sigma_{\rm i}$ and the coordinates of the center of the ${\rm i}$-th sphere, ${\bf r}_{\rm i}$, match (respectively) the instantaneous gyration radius, $R_{\rm g, i}$, and position of the center-of-mass (COM), ${\bf R}_{\rm cm, i}$, of the underlying PS subchain. 

The connectivity of soft-sphere chains is described~\cite{Vettorel,Guojie1} using bond, $V_{\rm b}$, and angular, $V_{\rm \theta}$, potentials, defined as: 
\begin{eqnarray}
\beta V_{\rm b}(d) &=& \frac{3d^{2}}{2b_{\rm BL}^2}, \\
\beta V_{\rm \theta}(\theta) &=& \frac{1}{2}k_{\rm BL}(1+\cos \theta),
\label{eq:vbonded}
\end{eqnarray}

\noindent
where $d$ and $\theta$ are the distance and angle between consecutive spheres and bonds in a soft-sphere chain, respectively. The parameters $b_{\rm BL}$ and $k_{\rm BL}$ control the 
strength of the potentials; their energy scale is expressed in units of $k_{\rm B}T$ ($\beta = 1/k_{\rm B}T$). 

The fluctuations of the radius, $\sigma$, of a sphere are controlled by a ``self-interaction'' potential, $V_{\rm s}(\sigma)$, similar to the Flory free energy~\cite{Flory}. 
This potential is defined as:
\begin{eqnarray}
\beta V_{\rm s}(\sigma) =  c_{\rm 1}\frac{\sigma^2}{N_{\rm b}} + c_{\rm 2}\frac{N_{\rm b}^2}{\sigma^{3}} 
\label{eq:Vsp}
\end{eqnarray}

The first term in Eq.~(\ref{eq:Vsp}) is of entropic origin and balances the swelling induced by the second term, which implicitly accounts for binary intramolecular interactions between the $N_{\rm b}$ superatoms underlying the blob. The reader may notice that in previous studies~\cite{Vettorel, Guojie1, Guojie2, Guojie3} the interactions controlling the fluctuations of $\sigma$ were augmented by a term, proportional to $1/\sigma^6$ (following Lhuillier~\cite{Lhuillier}). In the first implementation of the soft-sphere model~\cite{Vettorel} the $1/\sigma^6$ 
contribution enabled the description of a broad range of concentration regimes, e.g. dilute solutions. However, in melts the $1/\sigma^6$ term is significantly smaller than 
the $1/\sigma^{3}$ term~\cite{foot1} and thus can be omitted. 

The effective non-bonded interactions between two spheres, ${\rm i}$ and ${\rm j}$, are described by a Gaussian potential:
\begin{eqnarray}
\beta V_{\rm nb}(r_{\rm ij}) = \varepsilon N_{\rm b}^2 \left(\frac{3}{2 \pi \bar{\sigma}^2}\right)^{3/2} \exp\left( - \frac{3 r_{\rm ij}^2}{2 \bar{\sigma}^2}\right)
\label{eq:vnonbonded}
\end{eqnarray}

\noindent
where $r_{\rm ij}$ is the distance between the centers of the spheres, $\bar{\sigma}^2 = \sigma_{\rm i}^2 + \sigma_{\rm j}^2$, and $\varepsilon$
controls the strength of repulsion~\cite{foot2}. $\beta V_{\rm nb}(r_{\rm ij})$ is obtained by approximating the interactions of superatoms underlying different
spheres by binary repulsive collisions and is proportional to the overlap of two Gaussian clouds. Each of these clouds describes the average distribution in space of superatoms with respect to the COM of the subchain they belong to. 

The soft-sphere model proposed for the PS is motivated by arguments~\cite{Suter,Vettorel,Guojie2, Guojie3} based on general polymer physics. 
The model aims to capture {\it long-wavelength} conformational and structural properties of PS melts, accurately enough for performing backmapping at one state point. 
In this work, we define state points using the density of the blobs $n N_{\rm BL}/V$ ($n$ and $V$ are, respectively, the number of chains and volume of the sample) 
and temperature $T$. 
Due to the simple Gaussian-potential approximation used for the non-bonded interactions, the soft-sphere model is not expected to be thermodynamically consistent, e.g. it will not reproduce the equation-of-state of the PS. For this purpose, more elaborated blob-based models can be developed using techniques such as the integral equation theory~\cite{Guenza_2012a,Guenza_2012b, QWang}  and others.\cite{Briels,Karn2015}

To sample the configurational space of PS melts described with the soft-sphere model, we benefit from an efficient Monte Carlo (MC) algorithm based 
on three types of moves: i) random change of sphere size, ii) random sphere displacement, and iii) slithering snake (reptation). A detailed presentation of the method 
can be found elsewhere.~\cite{Guojie1} Here, we summarize that the computational efficiency stems from a special particle-to-mesh calculation of nonbonded interactions, which avoids 
neighbor lists.~\cite{Guojie1, Guojie2}

\subsubsection{Parameterization strategy}
\label{sec:ParamPS}

The first step is to set the resolution of the soft-sphere model by choosing $N_{\rm b}$. Currently there are no rigorous rules for choosing $N_{\rm b}$, apart from that $N_{\rm b}$ must be ``large enough'' for the soft-sphere model to be valid but smaller 
than the entanglement length, $N_{\rm e}$. The last requirement is important for backmapping. When soft-sphere chains are substituted by the moderately CG polymers, 
the liquid structure and chain conformations in the melt must be relaxed on scales smaller or comparable to the average blob diameter. 
The condition $N_{\rm b} < N_{\rm e}$ warrants that this relaxation is achieved through a fast Rouse-like dynamics of short subchains and is not affected by surrounding topological constraints.~\cite{Guojie2} Had we employed a hierarchy with several blob-based models~\cite{Guojie2}, the condition $N_{\rm b} < N_{\rm e}$ would apply only to the last, highest resolution, blob-based model (where the reinsertion of 
the moderately CG description is performed). 
Depending on the method used to extract the entanglement length, simulations of PS have reported $N_{\rm e}$ in the range of $100-200$ monomers, which is equivalent to $200-400$ superatoms.~\cite{HarmandRev} These results agree with experiments.~\cite{Fleisher,Antonietti,Fetters} Therefore, we explore different resolutions satisfying the condition $N_{\rm b}\le 200$. 
The entanglement time in the  moderately CG model is $\tau_{\rm e} \simeq 4 \times 10^4 \tau$.

For each trial $N_{\rm b}$, the parameters $c_1,c_2,b_{\rm BL},k_{\rm BL}$, and $\varepsilon$ are determined using typical structural-based (or Inverse Boltzmann) coarse-graining.~\cite{Lyubartsev, Kremer2002, Peter} 
The parameters are chosen so that the probability distributions $P_{\rm BL}(\sigma)$, $P_{\rm BL}(d)$, and $P_{\rm BL}(\theta)$, 
as well as the pair-correlation function, $g_{\rm BL}(r)$, of the centers of the spheres in blob-based PS melts, reproduce closely their counterparts 
in reference samples described by the moderately CG model. These reference samples contain $50$ chains with $N_{\rm CG} = 960$ superatoms and 
were equilibrated through a variant~\cite{HarmandPS} of the configuration-assembly method proposed by Auhl et al~\cite{Auhl} and long CG molecular dynamics simulations. 
The density of chains in the reference samples is $n/V = 1.17 \times 10^{-2}$ chains/nm$^3$ and the temperature is set to $T=463$\;K. In the moderately CG reference samples, the polymers are partitioned into subchains with $N_{\rm b}$ superatoms. 
The length of the chains in the reference samples restricts the $N_{\rm b}$ that can be considered here to a limited set of values, to have an integer number of subchains, i.e. under the condition $N_{\rm b}\le 200$, we try $N_{\rm b} = 192$, $160$, $120$, and so on. 
After the subchains are identified, we calculate the distributions of: i) gyration radii of subchains, $P_{\rm ref, BL}(R_{\rm g})$; ii) distance between the COM's of subchains, sequential in the same PS molecule, $P_{\rm ref, BL}(d)$; and iii) angles between two vectors joining the COM of a subchain with the COMs of the preceding and succeeding subchain, $P_{\rm ref, BL}(\theta)$ (cf. Fig.~\ref{fig:hierarchical_modeling}(c)). The pair distribution function of the COMs of subchains, $g_{\rm ref, BL}(r)$, serves as the counterpart of $g_{\rm BL}(r)$. 

With the reference distributions in hand, the parameters of the soft-sphere model are optimized through an iterative procedure which simultaneously minimizes four merit functions, defined~\cite{PlatheTarget} as:
\begin{eqnarray}
\Delta_{f(x)} = \int_{\rm 0}^{x_{\rm cutoff}} w(x)\left[f(x)-f_{\mathrm{ref}}(x)\right]^2 dx,
\label{eq:target}
\end{eqnarray}

\noindent
where $f(x)=P_{\rm BL}(\sigma),P_{\rm BL}(d),P_{\rm BL}(\theta)$ and $g_{\rm BL}(r)$; $f_{\rm{ref}}(x)$ is the corresponding reference function. 
In this study, we set the weighting function $w(x)$ in Eq.(\ref{eq:target}) to a constant, $w(x) = 1/x_{\rm cutoff}$, so that it has no effect on $\Delta_{f(x)}$. 
For $P_{\rm BL}(\theta)$ no cutoff is required, since $\theta \in [0,\pi]$. For $f(x)=P_{\rm BL}(\sigma)$
and $P_{\rm BL}(d)$, the $x_{\rm cutoff}$ is chosen such that $f(x_{\rm cutoff})\simeq 10^{\rm -4}$. We use $x_{\rm cutoff} = 12\;\sigma_{\rm CG}$ for $f(x)=g_{\rm BL}(r)$, 
because $g_{\rm ref, BL}(r=12\;\sigma_{\rm CG})$ saturates, within the noise of the data, to unity.  
\begin{figure}[ht]
\includegraphics[width=0.35\textwidth]{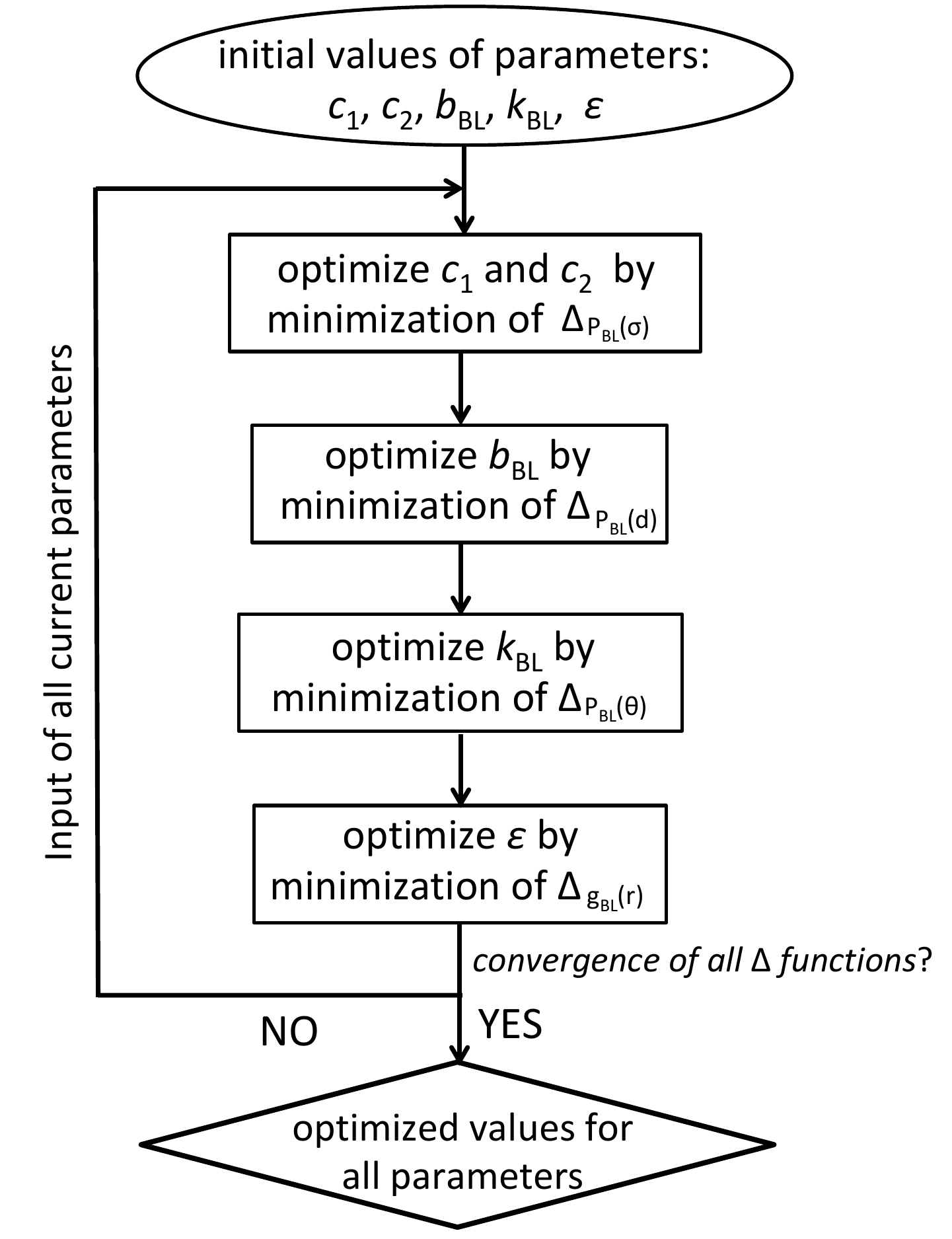}
\caption{Procedure used to parameterize the blob-based model of polystyrene melts.}
\label{fig:parameterization}
\end{figure}

To start the iterative procedure, the initial values of $b_{\rm BL}$ and $k_{\rm BL}$ are obtained by fitting 
the Boltzmann distributions of the potentials $V_{\rm b}(d)$ and $V_{\rm \theta}(\theta)$ from Eq.~(\ref{eq:vbonded}) 
to the reference distributions $P_{\rm ref, BL}(d)$ and $P_{\rm ref, BL}(\theta)$, respectively. The parameters 
$c_1$ and $c_2$ are obtained in a similar way by fitting the Boltzmann distribution of the potential 
$V_{\rm s}(\sigma)$, Eq.~(\ref{eq:Vsp}), to $P_{\rm ref, BL}(R_{\rm g})$. We emphasize that this first 
estimate of the parameters is approximate, because it assumes that the distributions are not correlated. A first-guess value for $\varepsilon$ 
can be obtained from $c_{\rm 2}$. To first order, the two parameters can be related to each other by equating the potential $c_{\rm 2} N_{\rm b}^2/\sigma^3$ (cf. Eq.~(\ref{eq:Vsp})), 
which captures the effect of the repulsion between intramolecular superatoms, to the ``self-interaction'' of a Gaussian density distribution. 
The latter is given by $\beta V_{\rm nb}(0)/2$; the prefactor $1/2$ avoids double counting. This approximation leads to $c_{\rm 2} = (\varepsilon/2)(3/4\pi)^{3/2}$. 

To perform the first iteration step, cf. Fig.~\ref{fig:parameterization}, we consider the soft-sphere model with the initial values of parameters. 
The number of blobs in the soft-sphere chains equals the number of subchains used to partition the moderately CG molecules in the reference samples.
The melt has $n=100$ soft-sphere chains and the volume is chosen such that $n/V$ 
matches the chain density in the reference systems. The temperature does not appear explicitly in the soft-sphere 
model, because all interactions in Eqs.~(\ref{eq:vbonded}) and (\ref{eq:Vsp}) are scaled by the thermal 
energy. The melt is equilibrated using the particle-to-mesh MC and the distribution function $P_{\rm BL}(\sigma)$ is 
extracted to calculate $\Delta_{P_{\rm BL}(\sigma)}$. The parameters $c_1$ and $c_2$ are increased or decreased by $5\%$ and a new 
simulation is performed. The procedure is repeated until $\Delta_{P_{\rm BL}(\sigma)}$ 
reaches a minimum value. The configurations from the simulation 
with the last optimized $c_1$ and $c_2$ are used to calculate $\Delta_{P_{\rm BL}(d)}$. The parameter $b_{\rm BL}$ is modified until $\Delta_{P_{\rm BL}(d)}$ reaches minimum. The same procedure is repeated for $k_{\rm BL}$ and $\varepsilon$; the corresponding merit functions are $\Delta_{P_{\rm BL}(\theta)}$ 
and $\Delta_{g_{\rm BL}(r)}$, cf. Fig.~\ref{fig:parameterization}. After updating all parameters, we commence a new iteration step. 
We recalculate $\Delta_{P_{\rm BL}(\sigma)}$ and improve $c_1$ and $c_2$ by minimizing again $\Delta_{P_{\rm BL}(\sigma)}$. Subsequently, 
the remaining parameters are improved, as in the first iteration step. Iteration steps are repeated until the minimum values of all 
merit functions, $\Delta_{f(x)}$, saturate. 

Through exploratory studies we find that the simple soft-sphere model can be parameterized to reproduce static properties of the reference PS melt when $N_{\rm b}\gtrsim 120$. Backmapping a soft-sphere model with the smallest possible $N_{\rm b}$ is preferable, because this procedure requires shorter relaxation times of reinserted microscopic details. Therefore, in this study we work with $N_{\rm b} = 120$.

\subsection{Validation of the soft-sphere model}
\label{sec:validation}
\begin{figure}[ht]
  \includegraphics[width=0.48\textwidth]{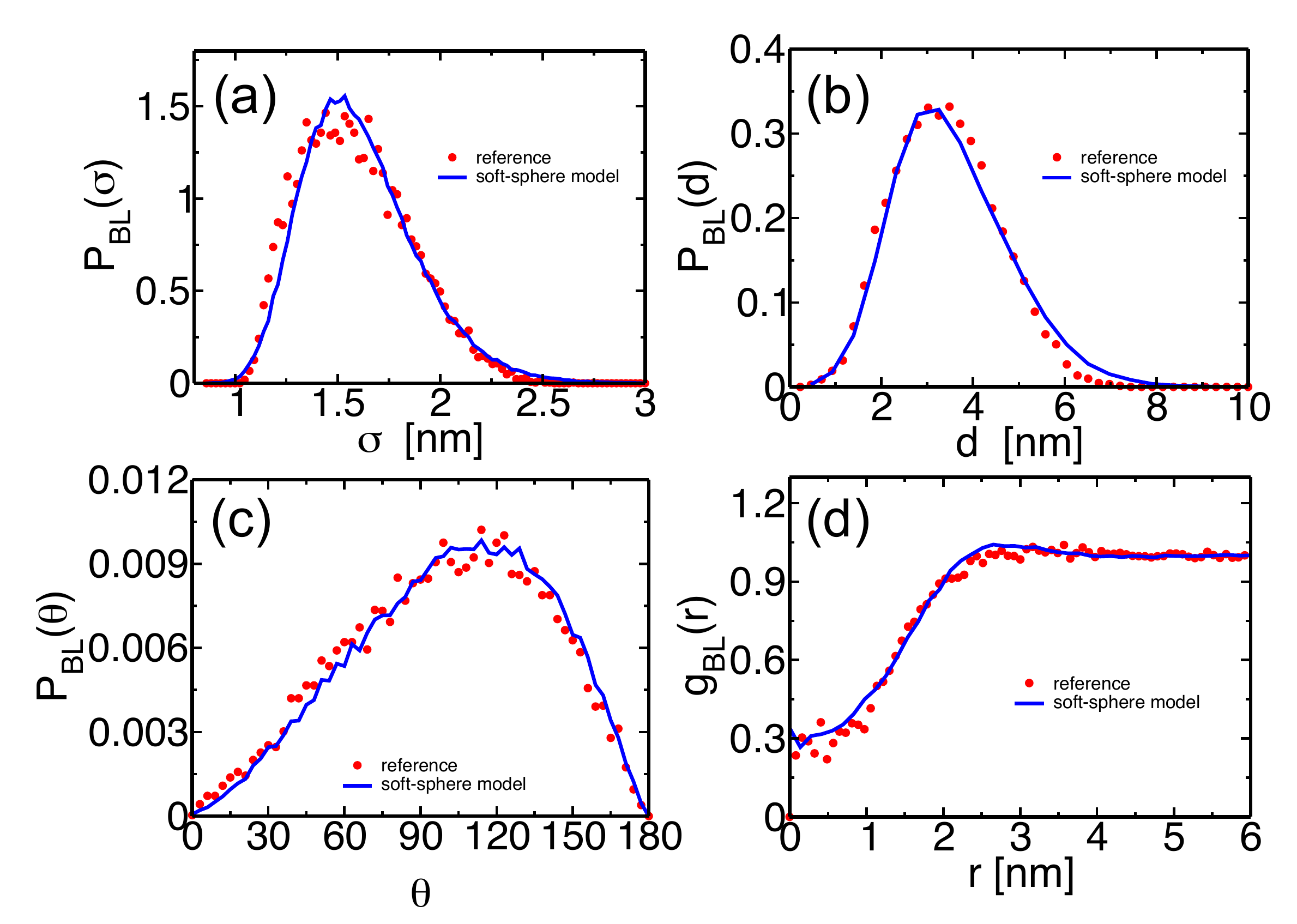}
	\caption{For a PS melt of soft-sphere chains corresponding to $N_{\rm b} =  120$, the lines show the probability distributions of a) sphere radius, $\sigma$,  
	b) bond length, $d$, between spheres, and c) angle, $\theta$, between successive bonds. The reference distributions calculated in moderately 
	CG melts of PS are plotted with symbols. d) For the same melt, the pair distribution function $g_{\rm BL}(r)$ of the centers of the soft spheres (lines) is compared with the pair distribution of the COMs of the subchains with $N_{\rm b} = 120$ superatoms (symbols). 
           }
\label{fig:parameter1}
\end{figure}

We illustrate that the melts equilibrated with the soft-sphere model indeed describe structual and conformational properties
of PS melts on length scales comparable or larger than the size of the blobs. For $N_{\rm b} = 120$, the four panels of Fig.~\ref{fig:parameter1} present the 
distributions $P_{\rm BL}(\sigma)$, $P_{\rm BL}(d)$, and $P_{\rm BL}(\theta)$, as well as the pair distribution function, $g_{\rm BL}(r)$, in a melt of soft-sphere chains (lines) 
and reference samples (symbols). The agreement of the curves is remarkable.  As will be demonstrated in the following, the small deviations observed in the plots 
do not propagate into the properties of backmapped PS samples.

\begin{figure}[ht]
\includegraphics[width=0.48\textwidth]{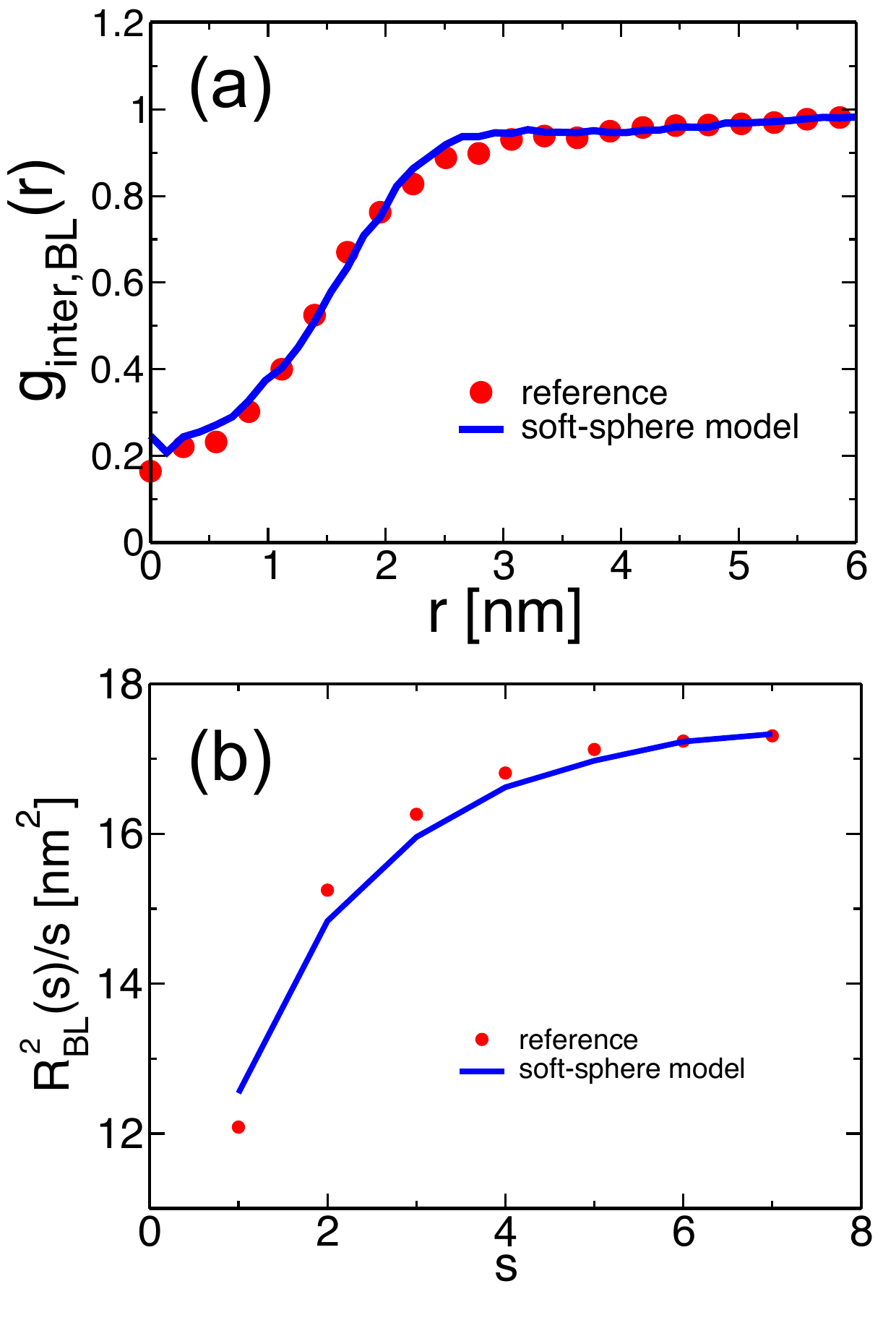}
\caption{(a) The line shows the intermolecular part, $g_{\rm inter,BL}(r)$, of the pair distribution function of the centers of the blobs 
	in PS melts of soft-sphere chains, corresponding to $N_{\rm b} =  120$. Symbols present the $g_{\rm inter,BL}(r)$ of COMs of subchains with $N_{\rm b} =  120$ 
	superatoms, calculated from the reference samples of moderately CG melts of PS. (b) A normalized internal distance plot, $R_{\rm BL}^2(s)/s$, 
	calculated in PS melts of soft-sphere chains corresponding to $N_{\rm b} =  120$ is shown with a line. Symbols present the internal distance plot calculated for the COMs 
	of subchains with $N_{\rm b} = 120$ superatoms in the reference samples.}
					        \label{fig:blob_conformations}
\end{figure}

Figs.~\ref{fig:blob_conformations}(a) and (b) quantify the accuracy of the soft-sphere model on scales comparable to the size of the
entire PS chain. Fig.~\ref{fig:blob_conformations}(a) presents for the soft-sphere model (lines) and reference samples (symbols) the intermolecular
part $g_{\rm inter, BL}(r)$ of the pair distribution functions plotted in Fig.~\ref{fig:parameter1}(d). In both curves, the correlation holes
of subchains  (evident depletion at small distances) and of entire PS chains (shallow depletion towards tail) can be identified. 
The two $g_{\rm inter, BL}(r)$ match each other closely, even in their long tails, demonstrating that the blob-based model captures 
the melt structure on large scales. Fig.~\ref{fig:blob_conformations}(b), compares polymer conformations in melts described by the soft-sphere (lines) 
and moderately CG  (symbols) models. We employ the internal distance plot, $R_{\rm BL}^2(s)/s$, which constitutes a very sensitive quantifier of equilibration.~\cite{Auhl} 
For the soft-sphere model, $R_{\rm BL}^2(s)$ is the mean-square distance between the centers of spheres in the same chain. In moderately CG melts, $R_{\rm BL}^2(s)$
is the mean-square distance between the COMs of subchains with $N_{\rm b} =  120$ superatoms, belonging to the same molecule. In both cases, $s$ is the
difference of ranking numbers of spheres (subchains) along chain contour. For large $s$ the plots follow each other closely; for blobs, 
located near each other along the chain, the relative deviation is at most $3.7\%$. The internal distance plot demonstrates the accuracy of the soft-sphere model in describing 
conformational properties on the scale of entire chains.

\section{Hierarchical backmapping strategy}

\subsection{Systems studied}
\label{sec:systems}

The hierarchical scheme described in the previous section can be applied to systems of any molecular weight for given density. In practice, for fixed pressure and temperature, the density of the polystyrene becomes chain-length 
independent for polymerization degrees larger than about 50-100 monomers.~\cite{HarmandRev} 
Therefore we expect that the soft-sphere model, developed in the previous section, is transferable to PS melts with arbitrary long chains. However, for the purposes of method development, we set here 
as a goal the equilibration of atomistic PS melts comprised of  $n=100$ chains with $N=480$ repeat units, that corresponds to a molecular weight of $M=50kDa$. These molecules 
are equivalent to moderately CG PS molecules with $N_{\rm CG} = 960$ superatoms and soft-sphere chains with $N_{\rm BL} = 8$ 
blobs each. These are the chain lengths for which the soft-sphere model is parameterized and for which reference 
samples are available~\cite{HarmandPS} (cf. Sec~\ref{sec:ParamPS}). Having the same chain lengths in backmapped and 
reference samples simplifies significantly the validation of the equilibration. The temperature in the backmapped and reference melts is $T=463$\;K, 
which is a typical processing temperature for PS.    

\subsection{Backmapping the blob-based model to the chemically specific coarse-grained model}
\label{sec:blob2CGPS}

The reinsertion of the degrees of freedom of the chemically specific CG model into equilibrated 
soft-sphere PS melts is conceptually similar to the strategy developed earlier~\cite{Guojie2} for generic microscopic (bead-spring) models. 
Technically, however, the backmapping of the chemically specific CG model is more involved, due to the more complex CG force field. The technicalities of the different backmapping steps are presented below. 

{\it Introducing moderately CG PS molecules:} Every soft-sphere chain in the melt is replaced by a moderately CG PS molecule 
in a matching conformation. Specifically, the COM and the radius of gyration (squared) of each subschain with $N_{\rm b}$ 
superatoms in the reinserted molecule must match the COM and the radius (squared) of the blob which has the same ranking number 
in the soft-sphere chain. We generate each moderately CG PS molecule in the vicinity of the blob-based chain it must replace (the exact location is not critical). 
Setting the first bead of a generated molecule to A type, the remaining beads 
are added stepwise. A and B beads alternate. In this work, we assign randomly to each $B$ bead one of the four subtypes (cf. Sec.~\ref{model:CGPS}) 
to make PS chains atactic. The length, bond angle, and torsional angle of the bond connecting the added superatoms to the part of the molecule
that has been already constructed, are randomly drawn from Boltzmann distributions. Each distribution depends on the appropriate bonded potential. 
Once the PS molecules are generated, we associate~\cite{Guojie2} with each subchain two external potentials: 
$\beta V_{\rm cm, i} = k_{\rm cm}({\bf r}_{\rm i} - {\bf R}_{\rm cm, i})^2$ and $\beta V_{\rm g, i} =  k_{\rm g}(\sigma_{\rm i}^2 - {R}_{\rm g, i}^2)^2$. 
Here ${\bf R}_{\rm cm, i}$ and ${R}_{\rm g, i}^2$ are the COM and squared radius of gyration of $\rm i$-th reinserted subchain, respectively.
These potentials affect all superatoms in the $\rm i$-th subchain, because ${\bf R}_{\rm cm, i}$ and ${R}_{\rm g, i}$ depend on the coordinates 
of all these superatoms. The parameters controlling the strength of the external potentials are empirically set to $k_{\rm cm}=100\;k_{\rm B}T/\sigma_{\rm CG}^2$ 
and $k_{\rm g}=100\;k_{\rm B}T/\sigma_{\rm CG}^4$. With the external and bonded potentials simultaneously activated (non-bonded potentials are turned off), the configuration of generated moderately CG molecules is subjected to MD simulation. This simulation ``drives'' every molecule into the corresponding soft-sphere chain. At this stage the molecules do not interact with each 
other, therefore the simulation takes negligible time. 

{\it Recovering microscopic excluded volume:} After the soft-sphere chains are replaced by the moderately CG PS molecules, we remove the external potentials and 
gradually activate~\cite{Guojie2} the non-bonded interactions between superatoms. For this purpose we use a ''push-off'' MD procedure~\cite{Auhl,Moreira} 
which removes overlaps between reiserted superatoms and recovers the original excluded volume characterizing the potentials $U_{\alpha,\beta}(r)$. We summarize here the main 
features of the push-off protocol; details are provided in the Appendix.

The potentials are ``force-capped'' according to the rule:
\begin{eqnarray}
	\label{eq:fc}
	U_{\alpha,\beta}^{\rm fc}(r) =
	 \left\{ \begin{array}{lll}
		(r-r_{\rm c(n)}) U_{\alpha,\beta}'(r_{\rm c(n)}) + U_{\alpha,\beta}(r_{\rm c(n)}),
		& r \le r_{\rm c(n)} \\
		\\ U_{\alpha,\beta}(r), \mbox{~~otherwise}
	\end{array}\right.
\end{eqnarray}

\noindent
Here $r_{\rm c(n)}$ is the force-capping radius and the original interactions are recovered when $r_{\rm c(n)} \rightarrow 0$. The push-off 
is accomplished in circles, each of them comprises a change in $r_{\rm c(n)}$ and local relaxation through a short MD simulation. 
For those beads that are not intramolecular $(1,5)$ neighbours, we decrease $r_{\rm c(n)}$ by a small step at the beginning of each cicle. For intramolecular $(1,5)$ neighbours, $r_{\rm c(1,5)}$ is adjusted in a special way in order to 
avoid significant distortions of polymer conformations during the push-off. To modify $r_{\rm c(1,5)}$ in the beginning 
of each cicle, we use an approach similar to the one developed for generic microscopic models~\cite{Moreira}. 
Namely, we quantify conformational distortions using the descriptor: 
\begin{eqnarray}
  I = \int_{n_{\rm 1}}^{n_{\rm 2}} \left[\frac{R_{\rm CG}^2(l)}{l} - \frac{R_{\rm CG,ref}^2(l)}{l} \right] {\rm d}l,
\label{eq:mastercurve1} 
\end{eqnarray}

\noindent
where $R_{\rm CG}^2(l)/l$ is the internal distance plot calculated in the PS melt after the previous circle of push-off 
is accomplished. $R_{\rm CG}^2(l)$ is the mean-square distance between superatoms in the same chain and $l$ is the
difference of ranking numbers of these superatoms along the chain backbone. The reference internal distance plot, $R_{\rm CG, ref}^2(l)/l$, 
is obtained from the reference samples of moderately CG PS melts. The boundaries in $t$ are empirically set to $n_{\rm 1} = 50$ and $n_{\rm 2} = 250$. 
We increase or decrease $r_{\rm c(1,5)}$, depending on whether $I$ is positive or negative. 

The MD push-off procedure lasts about $20\;\tau$ which is a neglible fraction of $\tau_{\rm e}$; about $0.05\%$. In practice, this run takes less than a day on 32 processors of a typical supercomputer.

{\it Local re-equilibration:} Once the excluded-volume interactions are recovered, the moderately CG PS melt is locally re-equilibrated through a standard MD simulation, 
which lasts about one $\tau_{\rm e}$. This is the most demanding computationally part of our procedure. This is the most demanding computationally part of our procedure. 
To give a feeling for the CPU resources that are typically required for re-equilibration, we mention that it takes about 32 days using the ESPResSo++ package~\cite{ESPResSo} 
(version 1.9.5) on 256 Xeon cores with frequency of 3.0 GHz. We observe that the local re-equilibration of PS melts for $\sim \tau_{\rm e}$, is about an order 
of magnitude slower -- in terms of the required CPU time -- comparing to melts described with generic microscopic bead-spring models.~\cite{Guojie2} The protracted 
computations are not related to the backmapping strategy 
 but are due to the complexity of the chemically-specific force-field. Because the largest relaxation time is determined by $\tau_{\rm e}$, the CPU time 
 required by our backmapping procedure {\it does not} depend on chain length and 
is proportional only to system volume.~\cite{Guojie2} In contrast, the relaxation time in brute-force MD simulations is proportional to 
the reptation time, given by $\tau_{\rm rep} \simeq \tau_{\rm e} (N/N_{\rm e})^3$; to make a simple scaling argument we use the approximate cubic 
power law of the initial reptation theory~\cite{Doi,GLAM}. The moderately CG PS melts considered here are only weakly entangled: 
based on the largest reported value $N_{\rm e} = 400$ (cf. Sec.~\ref{sec:ParamPS}) we obtain $N/N_{\rm e} \simeq 2.4$. The estimate for $\tau_{\rm rep}$ demonstrates 
that to equilibrate even this system, brute-force MD simulations would require about a year of continuous run (using the same amount of processors). 
Melts with slighly longer PS chains are entirely out of reach of brute-force MD simulations. 

\subsection{Validating backmapping of the moderately coarse-grained PS melts}

\begin{figure}[ht]
	\includegraphics[width=0.4\textwidth]{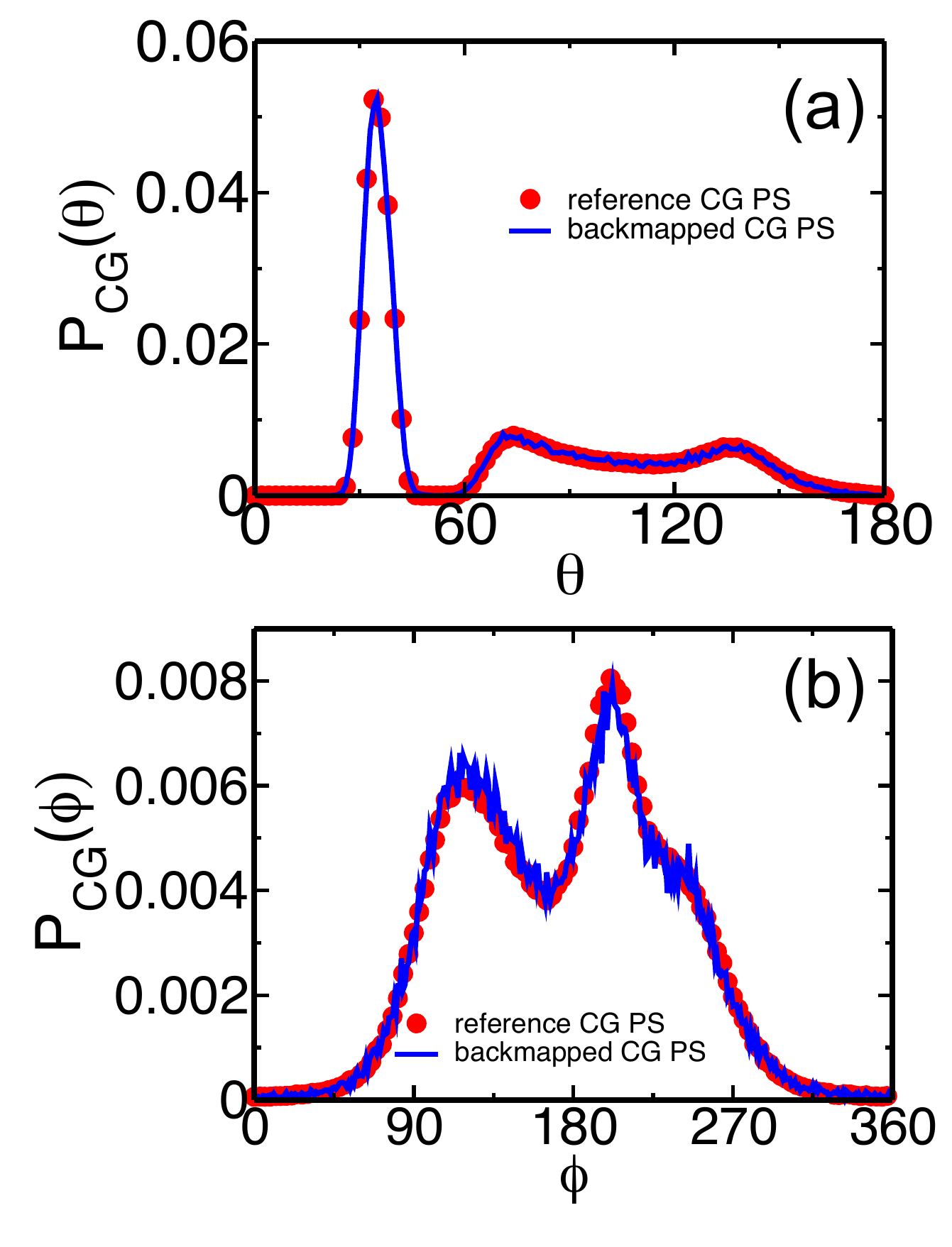}
	\caption{Comparison of probability distributions of (a) bond angles, $P_{\rm CG}(\theta)$ , and (b) torsional angles, $P_{\rm CG}(\phi)$, calculated in the backmapped (lines) 
	and reference~\cite{HarmandPS} (symbols) CG melts of PS.
}
	\label{fig:local_conf}
\end{figure}

We demonstrate the equilibration of backmapped moderately CG melts of PS by monitoring characteristic conformational and structural properties on local and global 
length scales. These properties are compared with their counterparts calculated in the reference samples. As a first simple test, we present in Figs. \ref{fig:local_conf}(a) and (b) 
the probability distributions of the bond angle and the torsional angle, $P_{\rm CG}(\theta)$ and $P_{\rm CG}(\phi)$. The data obtained for the backmapped melts (lines) 
are on top of the distributions (symbols) calculated in reference samples.

\begin{figure}[ht]
	\includegraphics[width=0.48\textwidth]{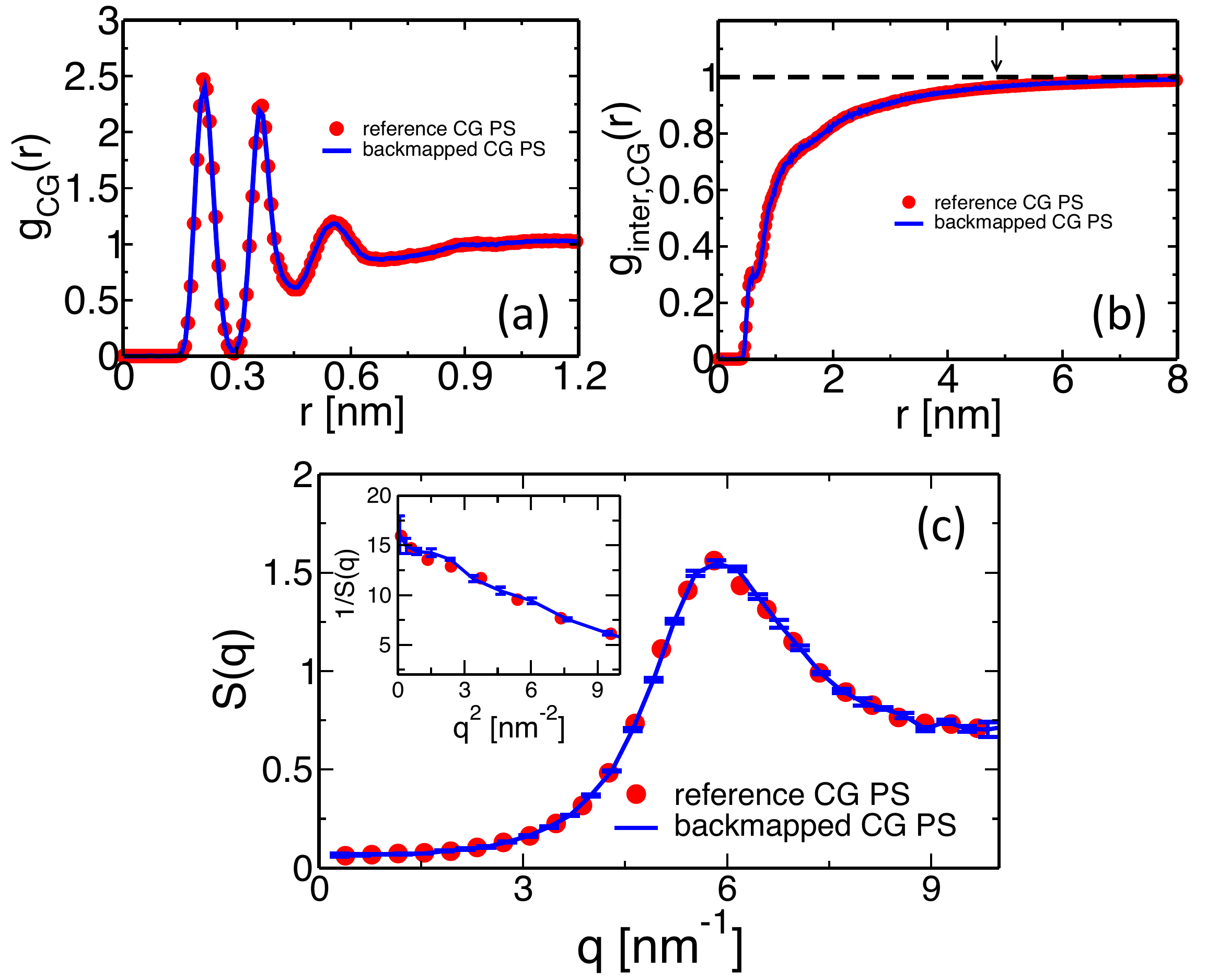}
	\caption{Comparison of (a) total, $g_{\rm CG}(r)$, and (b) intermolecular,  $g_{\rm inter, CG}(r)$, pair-distribution functions calculated in the backmapped (lines) 
	and reference~\cite{HarmandPS} (symbols) CG melts of PS. Panel (c) shows the structure factor of density fluctuations in the backmapped (lines) and reference (symbols) melts. The inset presents the behavior 
	of $1/S(q)$ for $q \rightarrow 0$, used to extract the isothermal compressibility.}
	\label{fig:gr_CGPS}
\end{figure}

The structure of the polymer liquid is quantified in Figs. \ref{fig:gr_CGPS}(a) and (b), presenting the total, $g_{\rm CG}(r)$, and the interchain, $g_{\rm inter, CG}(r)$, 
pair-distribution functions of superatoms in the backmapped (lines) and reference (symbols) samples. The plots obtained for backmapped and reference systems 
are indistinguishable from each other. The agreement between $g_{\rm inter, CG}(r)$ is particularly important -- thanks to the correlation-hole effect~\cite{deGennes} -- this quantity 
is a sensitive quantifier of the mesoscopic liquid structure on length scales comparable to the average size 
of the entire chain. The characteristic size of the correlation hole, defined by the average radius of gyration of the chains, is marked in Fig.~\ref{fig:gr_CGPS}(b) by the arrow. 
To illustrate more clearly that the backmapped melts reproduce correctly the long-wavelength density fluctuations of PS, we compare in Fig. \ref{fig:gr_CGPS}(c) 
their static structure factor $S(q)$ (line) with its counterpart (symbols) in reference samples. The two plots agree with each other within the statistical 
noise of the data. The latter is quantified via the error bars, corresponding to the standard deviation calculated from four independent backmapped CG melts of PS. 
From the data in Fig.~\ref{fig:gr_CGPS}, we obtain the isothermal compressibility, $\kappa_{\rm T}$, of atactic polystyrene at 463 K: $\kappa_{\rm T}=S(q \rightarrow 0)/\rho k_{\rm B}T\approx 8.7 \times 10^{-10} {\rm Pa}^{-1}$. Here $\rho$ denotes the number density of superatoms in the CG PS configurations. The $\kappa_{\rm T}$ characterizing the backmapped samples 
is remarkably close to $\kappa_{\rm T} = 8.4 \times 10^{-10} {\rm Pa}^{-1}$ that has been experimentally reported for PS at 473 K.\cite{polymer_handbook}

\begin{figure}[ht]
	\includegraphics[width=0.5\textwidth]{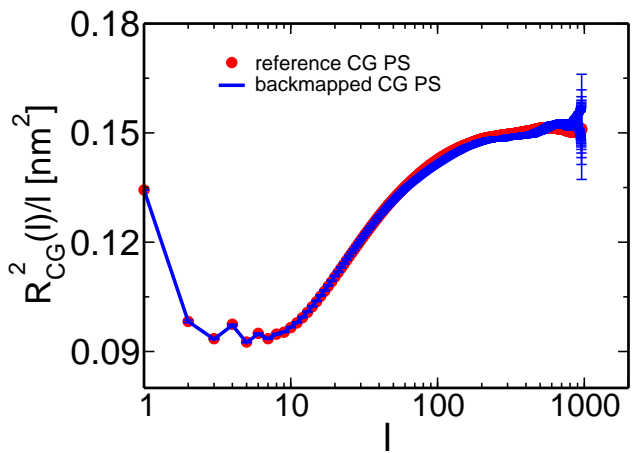}
	\caption{The line and the symbols present, respectively, the internal distance plots calculated in backmapped and reference~\cite{HarmandPS} CG melts of PS.}
	\label{fig:MSID_CGPS}
\end{figure}

One of the most stringent tests of equilibration is performed in Fig.~\ref{fig:MSID_CGPS}, where we present the internal distance plot $R_{\rm CG}^2(l)/l$ calculated in the backmapped (line) and reference (symbols) CG PS melts. The error-bars correspond to the $95\%$ confidence interval and, for clarity, 
are shown only for the plot obtained from the backmapped melts. The error-bars grow at the tail of the plot due to fewer data available for calculating $R_{\rm CG}^2(l)/l$. 
Within the noise of the data, the two plots match each other well and confirm the equilibration of the PS melts.

\subsection{Backmapping from the chemically specific coarse-grained model to the atomistic model}   
\label{CGPS2Atom}
In the following we describe the methodology for the backmapping from the moderately (chemically specific) CG scale to the atomistic one. Note that several different methods for re-introducing atomistic detail in CG polymer chain conformations have been appeared in the literature.~\cite{Tschoep2,VH2006,Peter,Carbone,Lobardi16} 

\begin{figure}[ht]
\includegraphics[width=0.4\textwidth]{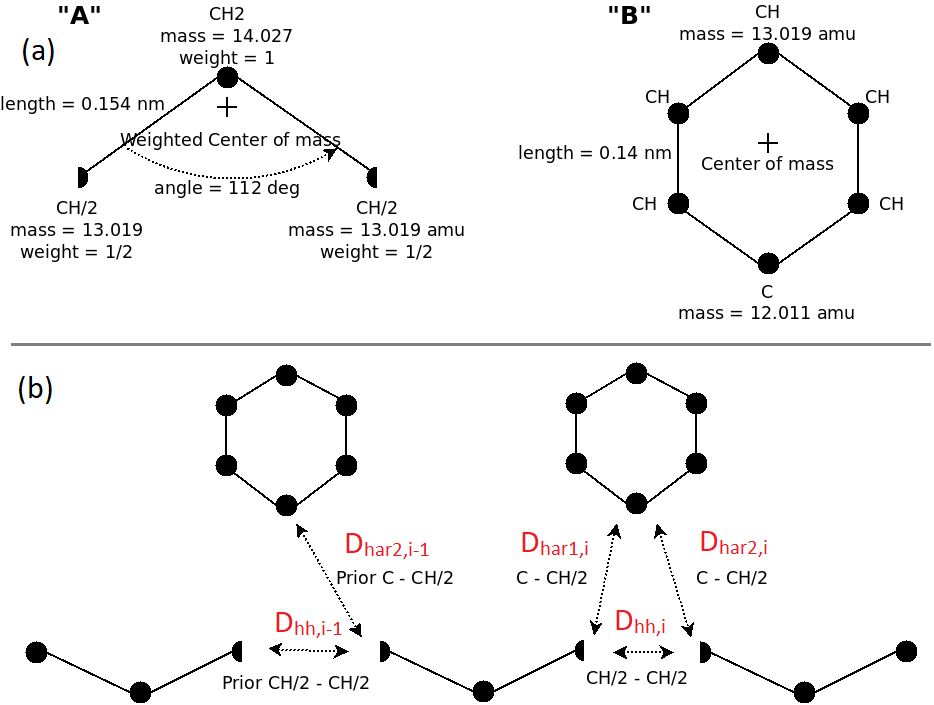}
\caption{(a) "Legos" that correspond to "A" and "B" CG groups. (b) Geometric re-introduction of the atomistic detail of two consequent PS monomers along the CG PS configurations.}
\label{fig:legos-min}
\end{figure}

Here, inspired from the above works, we propose a generic approach that is based on a "lego-like" construction of consecutive monomers along a macromolecular chain. The backmapping algorithm consists of the following stages:

\textit{Development of a CG "lego"-particle database:} 
Given a specific CG mapping scheme for PS chains, atomistic information about each CG particle is required. To obtain such information atomistic configurations of short PS chains are analyzed. 
More specifically, a particle database (class), based on the chosen CG mapping scheme, is constructed including two types of information: First, each member of the class represents a CG particle type and holds the atoms that define it, along with their relative positions, their masses and their contributing weights on the CG particle. In the specific CG PS model five different CG types are defined: One that describes the CG bead type "A" and four for the CG bead type "B" due to different tacticities.
The constructed A and B "legos" that correspond to the specific CG PS model are shown in Fig.~\ref{fig:legos-min}(a). Note that the lego A consists of a $CH_2$ and two half $CH$ groups, whereas lego B of the phenyl ring (five $CH$ and one $C$ groups). 

Second, from the analysis of the atomistic configurations, additional information is gathered about: (a) the average, and the distribution of distances between two consecutive CG particles ("AB" and "BA" CG bonds) and (b) the average and the distribution of angles between three ("ABA" and "BAB" CG angles).

\textit{Re-introduction of atomistic detail using a geometric algorithm:} 
In the next stage, given the CG PS configurations, obtained from the backmapping of the blob-based to the chemically specific CG representation, the atomistic ones are created.
This is achieved via a geometric algorithm, by putting each "lego" piece at the given center-of-mass of its corresponding CG particle in a consequent way. 
To avoid high overlaps the "legos" are rotated around their center-of-masses, to minimize distances between neighboring atoms; see Fig.~\ref{fig:legos-min}(b).

In more detail, for a specific PS monomer $i$ we define as $D_{hh,i}$ the distance between the two CH halves, which correspond to the same $CH$ united-atom group, and $D_{har1,i}$ and $D_{har2,i}$ the distances between the two halves and the "C" carbon of the aromatic ring, in the ith monomer.

In addition we define as: $\theta_{A,i}$=[$\theta_{A,x,i}$, $\theta_{A,y,i}$, $\theta_{A,z,i}$] and $\theta_{B,i}$=[$\theta_{B,x,i}$, $\theta_{B,y,i}$, $\theta_{B,z,i}$] the three (Eulerian) rotation angles, 
of the A and B CG particle-"lego" of the ith monomer, respectively. 
Based on the above we propose, along two consequent PS monomers, the following minimization problem:
\begin{equation}
\min_{\theta_{A,i},\theta_{B,i},\theta_{A,i+1}} F_{cost}
\end{equation}
where the cost function is defined as:
\begin{eqnarray}
&&F_{cost} = \nonumber\\
&&D_{hh,i}^2 + w_1((D_{har1,i}-D_{har})^2 + (D_{har2,i}-D_{har})^2) + \nonumber\\
 &&w_2(D_{hh,i-1}^2 + (D_{har2,i-1}-D_{har})^2) 
\label{eq:mastercurve2} 
\end{eqnarray}
where $D_{har}=0.151\;$nm is the (average) atomistic "C-backbone"  - "C-aromatic" bond length. $w_1$ and $w_2$ are weights with typical values $w_1$=$w_2$=0.7. 
Then, we minimize $F_{cost}$, over all possible angles $\theta_{A,i}$, $\theta_{B,i}$, and $\theta_{A,i+1}$ via a BFGS method. This is a very fast minimization procedure, since it involves information of only two consequent monomers.

The above procedure is repeated for all pairs of consequent monomers along the PS chain, in order to obtain a first realistic PS atomistic configuration.

\textit{Energy minimization:} 
The re-insertion of the atomistic detail on the CG PS configurations has been extensively tested and no strong overlaps between neighboring atoms were observed. However, due to possible correlations among the CG intramolecular (bonds, angles, dihedrals) degrees of freedom, as well as due to the fact that the geometric method does not involve thermal fluctuations, overlaps between atoms cannot be fully avoided. To treat such overlaps, in the third stage an energy minimization procedure for all atoms is required. 

We perform such a "global", with respect the number of atoms that are involved, minimization scheme in two steps: First, the non-bonded interaction potential, between all UA groups, is switched off and the bonded (involving all atomistic bonds, angles and dihedrals) potential is minimized through a steepest descent method, while the center-of-masses corresponding to the CG beads A and B are restrained in their original positions. 
Second, the non-bonded potential is switched on and the full atomistic force field (all bonded and non-bonded interactions) is minimized via a conjugated gradient algorithm.

\textit{Short MD runs:} 
Finally, in order to relax remaining stresses in the systems and to obtain a realistic atomistic trajectory at the appropriate temperature, a short NVT MD run, of about 100\;ps, is executed.

\subsubsection{Validating backmapping of atomistic, low MW, PS systems}

The quantitative validation of the backmapping from the chemically specific CG model to the atomistic one is a challenging issue. Indeed, a direct comparison of the atomistic configurations of high MW derived from the CG model with "reference" data is not possible, since in principle there are no accurate atomistic data for such polymer configurations. 

Therefore, in order to validate the backmapping procedure we decide to use a low MW  (1kDa, 10mer) PS melt, at T=463 K. 
First, we perform long atomistic MD simulations of atactic 10mer PS melts, using the model described in section~\ref{sec:models}. 
These simulations are performed in the NPT ensemble, whereas tail corrections for the energy and pressure were applied. The integration time step was 1 fs, whereas the overall atomistic simulation time of the production runs was 100ns.

Second, we apply the backmapping methodology, described in the previous section, to a CG 10 mer PS system.~\cite{HarmandPS} 
Thus, we obtain atomistic configurations of 10mer PS without performing atomistic MD simulations. 



\begin{figure}[ht]
\includegraphics[width=0.47\textwidth]{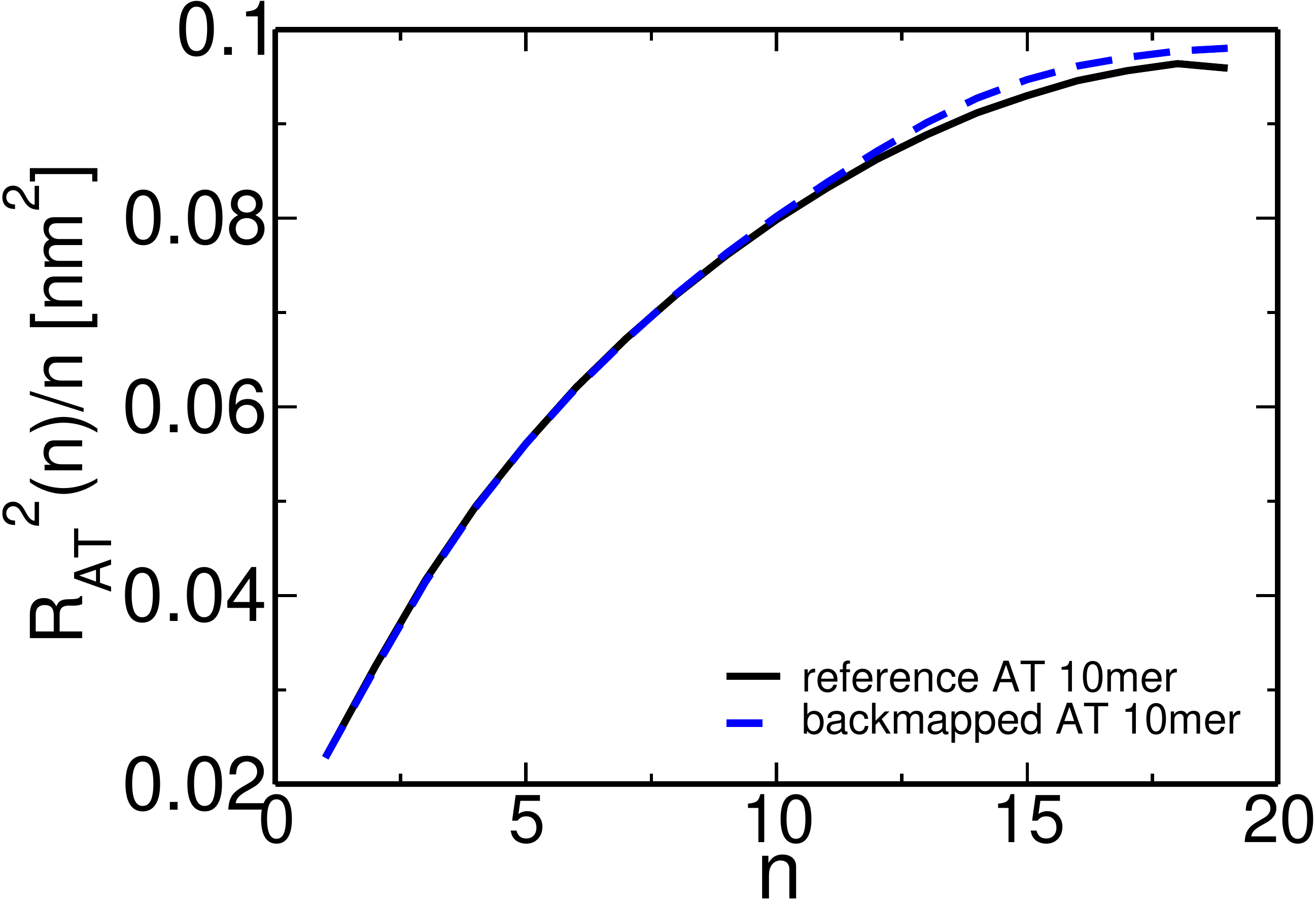}
\caption{Internal distance distribution, analyzed in the united-atom level, of a typical system (MW = 1 kDa) from long atomistic MD simulations and after backmapping of the CG configurations.}
\label{fig:intdist_10mer}
\end{figure}

Therefore, for the low MW PS melt atomistic data are gathered both from long atomistic MD simulations ("reference" data), and from CG systems after employing the CG-to-atomistic backmapping procedure.
As a direct "intramolecular" check of the backmapping process we examine the internal distance distribution, analyzed in the united-atom level, of a typical short-chain system (MW = 1 kDa) from long atomistic MD simulations and after backmapping of the CG configurations. 
In more detail, the $R_{\rm AT}^2(n)/n$, is calculated, where $R_{\rm AT}^2(n)$ is the average mean square distance between $n$ consequent backbone carbon atoms of the atomistic PS chains.
Data are shown in Fig.~\ref{fig:intdist_10mer}. As we can see the agreement between the two sets of data is excellent for all internal distances. 
Note that a similar agreement has been found before for short PS chains following a different back mapping approach.~\cite{Fritz}

\begin{figure}[ht]
\includegraphics[width=0.32\textwidth]{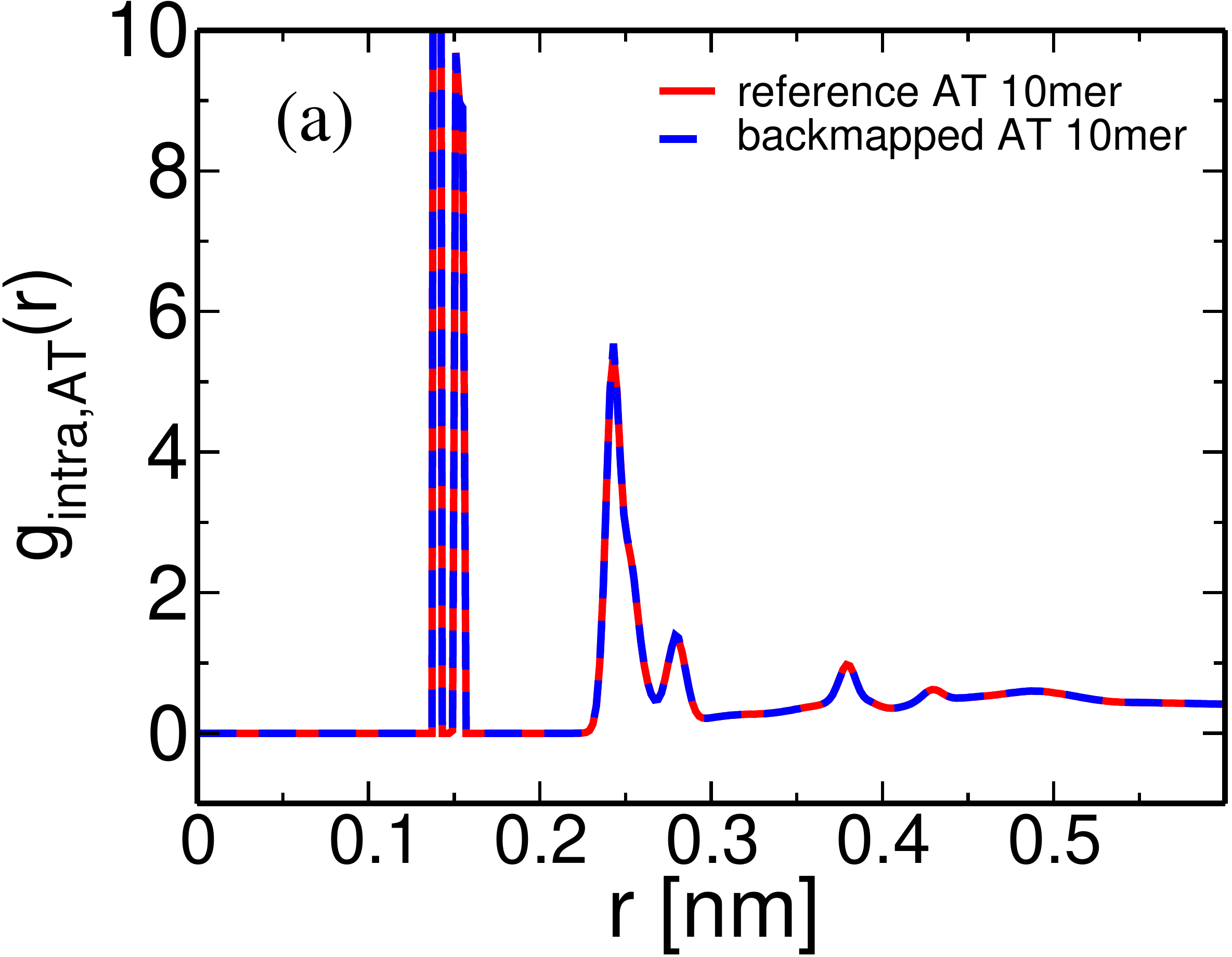}
\includegraphics[width=0.34\textwidth]{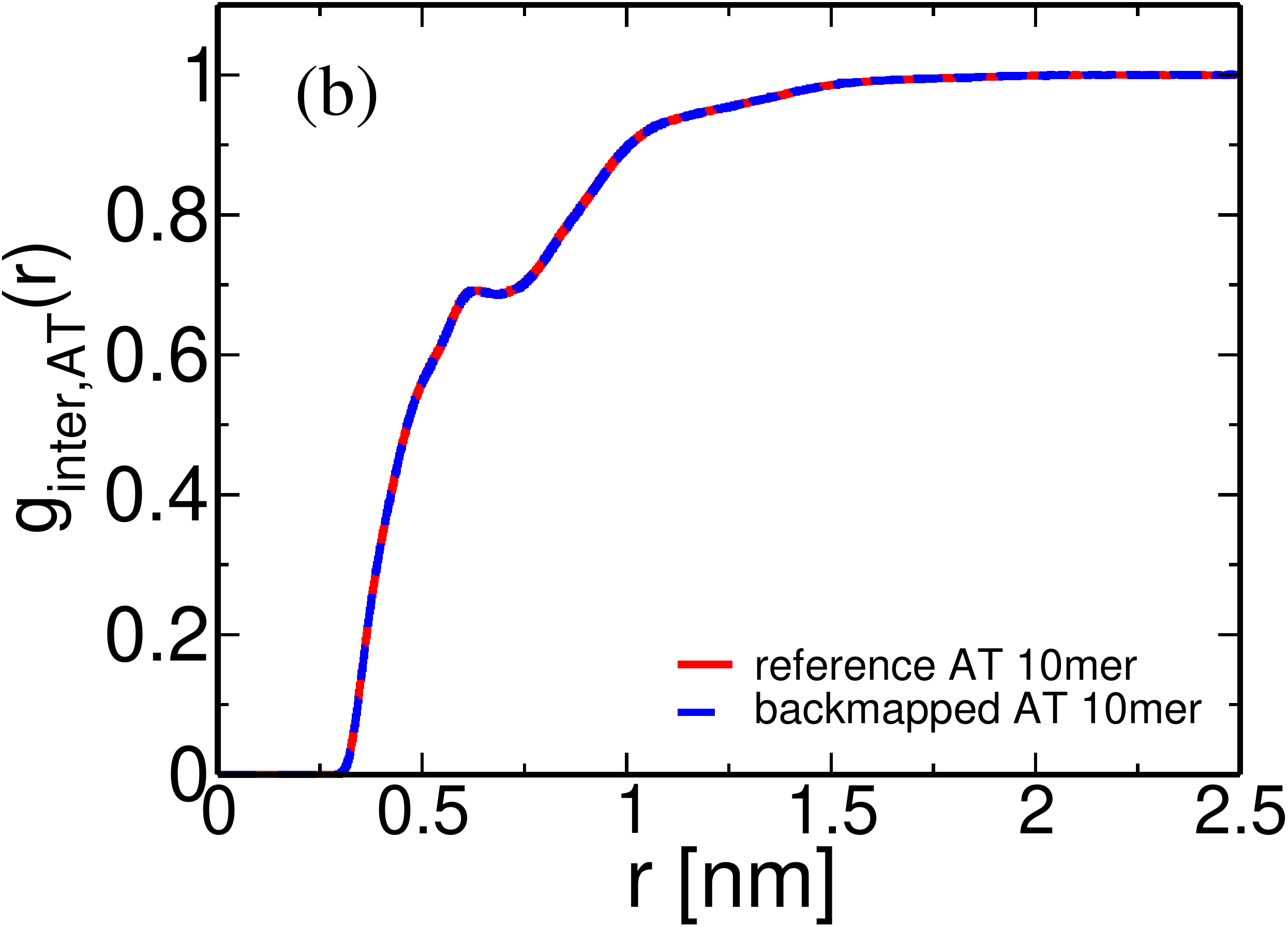}
\caption{(a) Intramolecular, and (b) Intermolecular pair distribution function for a PS melt (MW = 1 kDa, T =463 K), obtained from long atomistic MD simulations, and after backmapping of the CG configurations.}
\label{fig:gofr_10mer}
\end{figure}

The short chain PS atomistic configurations obtained from the CG ones through the backmapping procedure can be further examined by calculating the atomistic pair distribution functions, $g_{\rm AT}(r)$, and comparing them against the data derived directly from the long atomistic runs.
Results are presented in Figs.~\ref{fig:gofr_10mer}(a)(b). In both cases all correlations are included.
First, in Fig.~\ref{fig:gofr_10mer}(a) data for the intramolecular distribution function, $g_{\rm intra,AT}(r)$, are shown. 
The agreement between the two curves is excellent for all length scales providing thus a direct evidence of the applicability of the coarse-grained model to preserve the long length internal chain structure. 
Note, that these data include all intramolecular correlations, in contrast to the internal distance distribution data shown in Fig.~\ref{fig:intdist_10mer}, where only correlations along the backbone atoms are considered. Thus the agreement between the two sets of data shown in Fig.~\ref{fig:gofr_10mer}(a) proves the ability of the backmapping methodology to reproduce the full intramolecular structure.

The intermolecular pair distribution function, $g_{\rm inter,AT}(r)$, is shown in Fig.~\ref{fig:gofr_10mer}(b). The excellent agreement between the two different sets of data, clearly demonstrates that the local packing in the backmapped atactic PS melts is also well reproduced.
Note that the agreement between the intermolecular distribution functions based on different correlations, e.g. of "CH$_2$-CH$_2$" and of "phenyl-phenyl" groups (data not shown here) is of similar quality. 

\subsubsection{High MW atomistic PS melts}
After validating the "CG-to-atomistic" backmapping methodology we can proceed to applying this approach to the high MW (50kDa) CG PS melts, which have been obtained through the "blobs-to-CG" backmapping procedure described in Section~\ref{sec:blob2CGPS}.
By doing this, atomistic conformations of 480mer (MW=50 kDa) PS melts are directly obtained. Such well equilibrated high MW atomistic PS configurations cannot be generated through brute force MD simulations even with the most powerful computing resources.  

\begin{figure}[ht]
\includegraphics[width=0.4\textwidth]{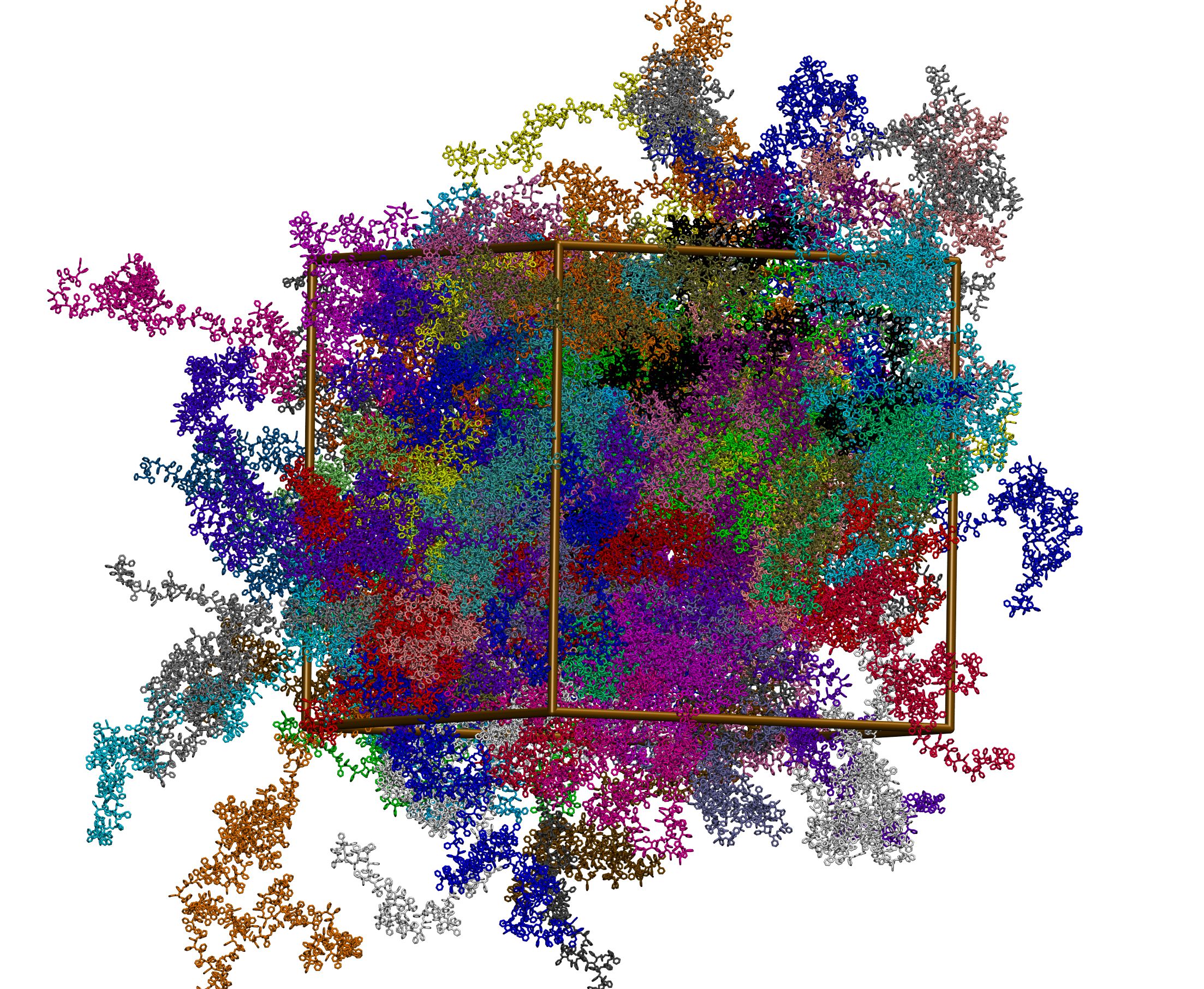}
\caption{Equilibrated atomistic configuration of a 50 kDa (480mer) PS melt with 100 chains (T =463 K). Chains are shown unwrapped (with the center-of-mass of each chain in the simulation box) and with different colors for clarity.} 
\label{fig:snap_atom}
\end{figure}

A typical well-equilibrated atomistic configuration (snapshot) of 50 kDa PS melts, with 100 chains is shown in Fig.~\ref{fig:snap_atom}.

\begin{figure}[ht]
\includegraphics[width=0.39\textwidth]{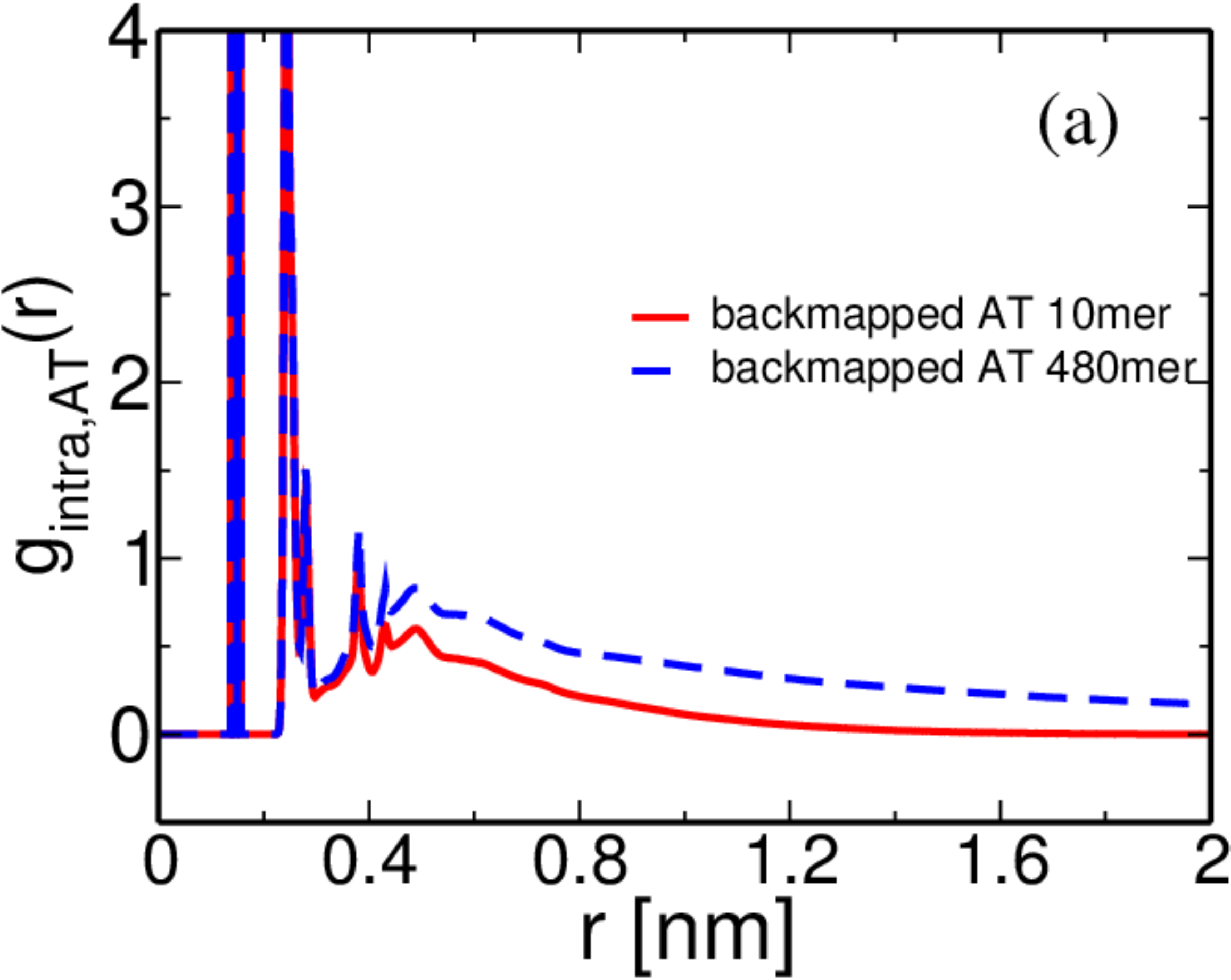}
\includegraphics[width=0.4\textwidth]{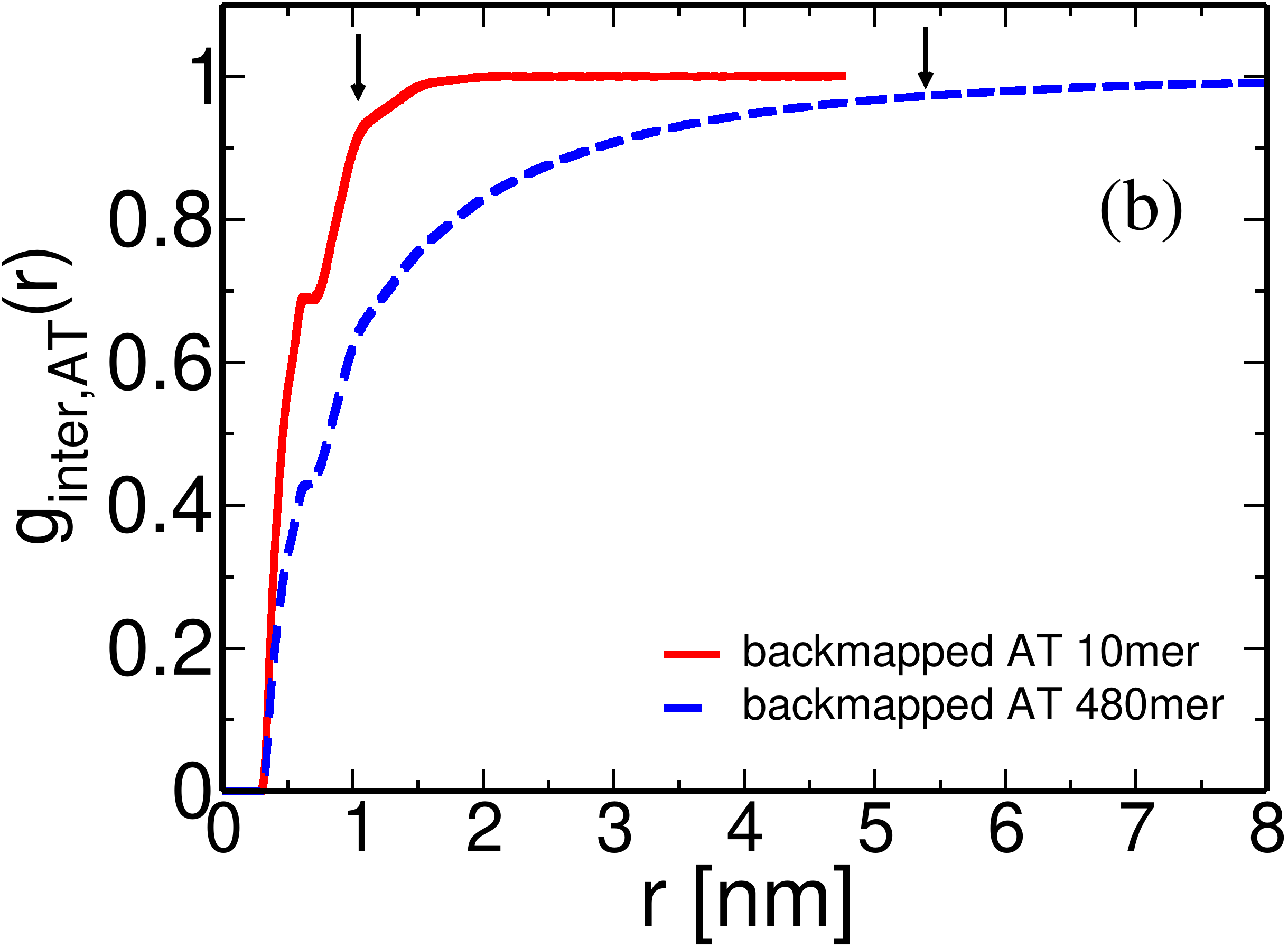}
\caption{(a) Intramolecular, and (b) Intermolecular, pair distribution functions for different PS melts, obtained directly from backmapping of the CG configurations. Arrows denote the average radius of gyration of the chains for both systems, $R_{g}(1 {\rm kDa})=0.9\;$nm and $R_{g}(50 {\rm kDa})=5.4\;$nm.}
\label{fig:gofr_480mer}
\end{figure}

The structure of polymeric chains, and its dependence on the molecular weight, is of special importance. Thus, in the following we compare the pair distribution functions of the long (entangled) PS chains with the data from the short (oligomeric) PS discussed above. 
The structure of the derived atomistic PS 480mer configurations is examined in Figs.~\ref{fig:gofr_480mer}(a) and (b).
In Fig.~\ref{fig:gofr_480mer}(a) data for the atomistic intramolecular distribution function, $g_{\rm intra,AT}(r)$ of both low (10mer) and high (480mer) molecular weight PS melts are shown.
The agreement between the two curves at very short distances of about $0.5\;$nm is expected. Indeed, correlations in such length scales involve neighboring atoms along the polymer chain (i.e. atoms belonging only in one-two consequent monomers), thus being very similar for the low and high molecular weight chains. 
For longer chains intramolecular distribution functions approach zero, as expected, however, $g_{\rm intra,AT}(r)$ for 480mers PS chains are extended to much longer lengths.

Different is the case for the overall chain packing, which is directly related with the correlation hole of the polymer chains.~\cite{deGennes} Data about the intermolecular pair distribution function, $g_{\rm inter,AT}(r)$, are shown in Fig.~\ref{fig:gofr_480mer}(b). 
It is obvious that the correlation hole extends over a distance of the order of the average radius of gyration of the chains; the latter is shown, for both systems, in Fig.~\ref{fig:gofr_480mer}(b) with arrows. 
In addition, for the high molecular weight chains the correlation hole becomes less deep, as has been also discussed before.~\cite{HarmandRev}

\begin{figure}[ht]
\includegraphics[width=0.45\textwidth]
{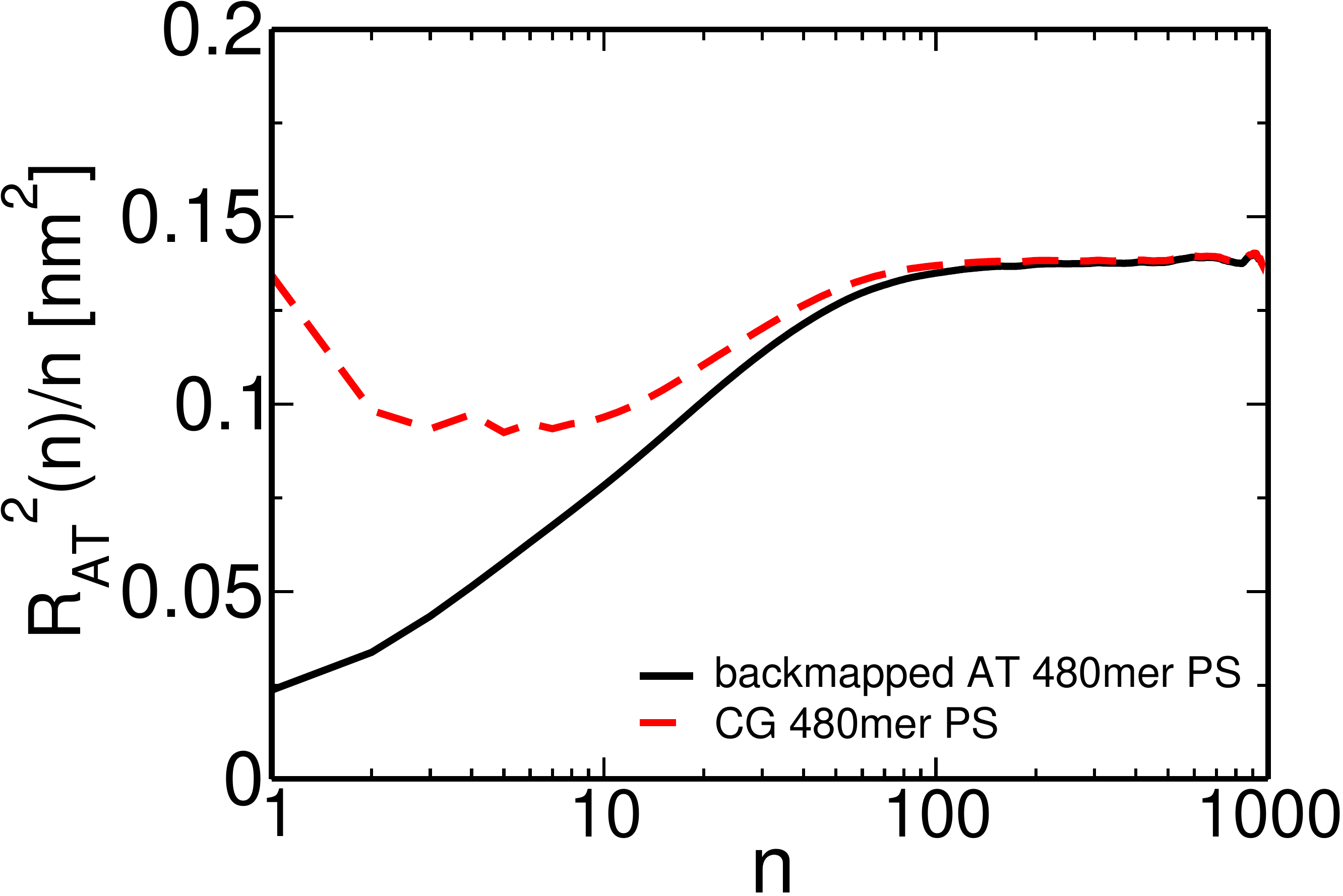}
\caption{Internal distance distribution, analyzed in the united-atom level, of a high MW PS chain obtained directly after backmapping of the CG configurations,without performing long atomistic MD runs.} 
\label{fig:intdist_480mer}
\end{figure}

We present in Fig.~\ref{fig:intdist_480mer} the internal distance plot $R_{\rm AT}^2(n)/n$ calculated in one representative backmapped sample of an atomistic 50kDa PS melt (solid black line). $R_{\rm AT}^2(n)$ stands for the average mean square distance between $n$ consequent backbone carbon atoms of the atomistic PS chains. 
For comparison, the internal distance plot calculated for the sample of the moderately CG melt used to backmap this atomistic sample is also shown (dashed red line).
For the atomistic sample, the internal distances start from a small value and then smoothly reach a plateau for $n$ around 80-100, which corresponds to 40-50 PS monomers.
This behavior at small $n$ differs from the internal distance plot in the moderately CG representation, where $R_{\rm AT}^2(n)/n$ appears non-monotonous and more structured. 
Such differences are expected because of the averaging of the monomeric degrees of freedom performed in the CG groups. 
In contrast, the two plots match each other at large $n$, demonstrating that our ``CG-to-atomistic'' backmapping procedure conserves
the global conformational properties that have been already equilibrated during the ``blob-to-CG'' backmapping stage. 


\section{Conclusions and Outlook}
Preparing equilibrated samples of highly entangled polymer melts serves as a first, but by no means trivial, step for studying their dynamical and rheological properties in and out of equilibrium using computer simulations. In this work, we propose a method for efficiently generating well-equilibrated configurations of high molecular-weight polymer melts, 
described with chemically-specific atomistic models. The method is build on a hierarchy of models, which describe the same material with different resolutions: 
from mesoscopic to atomistic. The modeling hierarchy includes a soft blob-based, a moderately coarse-grained, and an atomistic model. If required, 
more soft blob-based models can be incorporated, to accelerate equilibration of larger scales. Each model in the hierarchy is parameterized to reproduce key conformational 
and structural properties characterizing the melt, when described with the next finer-resolution model. First, the melt is efficiently equilibrated using the 
soft blob-based model. The details of the next level are reinserted, the system is re-equilibrated, and we proceed to the next reinsertion step. During each 
step of backmapping only local re-equilibration is required. Therefore, the computation time is independent of the molecular weight of the polymer that is considered.

Here, as a proof-of-concept, we use the method to equilibrate atomistically-resolved samples of melts of long attactic polystyrene chains. The preparation of 
such systems using brute force MD simulations is unfeasible, even with modern supercomputers. We emphasize that the method is not restricted to  polystyrene and 
can be straightforwardly applied to equilibrate melts of other polymeric materials, where moderately CG models exist.~\cite{Tschoep1,Hess,TakahKremer,Doxastakis,Faller,Rousseau}

\section*{Appendix: ``push-off'' procedure for backmapping the blob-based model to the chemically specific coarse-grained model}

The whole ``push-off'' procedure is accomplished in 100 cycles of a molecular dynamics simulation, where each cycle is comprised of $2 \times 10^4$ steps, 
and time step $\Delta t = 0.0001 \tau$ ($\tau$ is time unit in the CG PS model). (a) For all particle pairs ij other than the intramolecular (1,5) pairs $r_{\rm c(n)}$ 
is linearly decreased from $0.9 \mbox{~} r^{\rm cutoff}_{\rm ij}$ ($r^{\rm cutoff}_{\rm ij}$ is the cutoff of the LJ-like potential between i and j particles) to $0.5 \mbox{~} r^{\rm cutoff}_{\rm ij}$ within the first 80 cycles, after which particle overlaps for those pairs can be successfully removed. (b) For the intramolecular (1,5) pairs, the corresponding force-capped radius, $r_{\rm c(1,5)}$, is allowed to fluctuate (instead of decreasing linearly as in case (a)) to prevent large distortions of chain conformations. For this purpose, 
after each cycle, we quantify via $I$ (see Eq.(\ref{eq:mastercurve1} in Sec \ref{sec:systems}) the deviation of the internal distance plot from the reference one.
Before starting the next cycle we decrease (or increase) $r_{\rm c(1,5)}$ by $0.01\sigma_{1-5}$ if the calculated $I$ is negative (or positive). In this way, $r_{\rm c(1,5)}$ 
can fluctuate within the first 80 simulation cycles. In the last 20 cicles it is reduced linearly to $0.5 \mbox{~} r^{\rm cutoff}_{(1,5)}$.

\begin{acknowledgments}
The computing time granted by the John von Neumann Institute for Computing (NIC) on the supercomputer JURECA at J{\"u}lich Supercomputing Centre (JSC) is gratefully acknowledged. 
This work has been supported by the European Research Council (ERC) under the Seventh Framework Programme (FP7/2007-2013)/ERC Grant Agreement No. 340906-MOLPROCOMP.\\
\end{acknowledgments}



\begin{thebibliography}{999}
\bibitem{foot1} In previous implementations of the soft sphere model the prefactor $N_{\rm b}$ was adsorbed into $c_{\rm 2}$. Here it appears explicitly in the term $c_{\rm 2}N_{\rm b}^2/\sigma^{3}$ (see Eq.(\ref{eq:Vsp})) to highlight the relevance to the Flory free energy.
\bibitem{foot2} In previous implementations of the soft sphere model the prefactor $N_{\rm b}^2$ was adsorbed into $\varepsilon$.
\end{thebibliography}
\end{document}